\documentclass[manuscript, screen]{acmart}

\acmJournal{JACM}

\usepackage{verbatim}
\usepackage{graphicx}
\usepackage{epstopdf}

\usepackage{booktabs}
\usepackage{array}
\usepackage{multirow}
\usepackage{subfigure}
\usepackage{pdflscape}

\begin{document}

\title{Information flow based defensive chain for data leakage detection and prevention: a survey}

\author{Ning Xi}
\email{nxi@xidian.edu.cn}
\affiliation{%
  \institution{Department of Cyber Engineering, Xidian University}
  \streetaddress{Taibai Road, NO.2}
  \city{Xi'an}
  \state{Shaanxi}
  \country{China}
  \postcode{710071}
}

\author{Chao Chen}
\email{chao.chen@jcu.edu.au}
\affiliation{%
  \institution{College of Science and Engineering, James Cook University}
  \streetaddress{1 James Cook Dr, Douglas}
  \city{Townsville}
  \state{Queensland}
  \country{Australia}
 }
\authornote{Chao Chen is the corresponding author.}

\author{Jun Zhang}
\affiliation{%
  \institution{School of Software and Electrical Engineering,  Swinburne University of Technology}
  \streetaddress{John Street, Hawthorn}
  \city{Melbourne}
  \state{Victoria}
  \country{Australia}
 }

\author{Cong Sun}
\affiliation{%
  \institution{Department of Cyber Engineering, Xidian University}
  \city{Xi'an}
  \country{China}
}

\author{Shigang Liu}
\affiliation{%
  \institution{School of Software and Electrical Engineering,  Swinburne University of Technology}
  \city{Melbourne}
  \state{Victoria}
  \country{Australia}
}

\author{Pengbin Feng}
\affiliation{%
  \institution{Department of Cyber Engineering, Xidian University}
  \city{Xi'an}
  \country{China}
}

\author{Jianfeng Ma}
\affiliation{%
  \institution{Department of Cyber Engineering, Xidian University}
  \city{Xi'an}
  \country{China}
}



\renewcommand{\shortauthors}{Xi et al.}

\begin{abstract}
Mobile and IoT applications have greatly enriched our daily life by providing convenient and intelligent services. However, these smart applications have been a prime target of adversaries for stealing sensitive data. It poses a crucial threat to users' identity security, financial security, or even life security. Research communities and industries have proposed many Information Flow Control (IFC) techniques for data leakage detection and prevention, including secure modeling, type system, static analysis, dynamic analysis, \textit{etc}. According to the application's development life cycle, although most attacks are conducted during the application's execution phase, data leakage vulnerabilities have been introduced since the design phase. With a focus on lifecycle protection, this survey reviews the recent representative works adopted in different phases. We propose an information flow based defensive chain, which provides a new framework to systematically understand various IFC techniques for data leakage detection and prevention in Mobile and IoT applications. In line with the phases of the application life cycle, each reviewed work is comprehensively studied in terms of technique, performance, and limitation. Research challenges and future directions are also pointed out by consideration of the integrity of the defensive chain.  
\end{abstract}



\maketitle

\section{Introduction}
IT applications have been prevalent in the last ten years, especially with the development of mobile terminals, cloud computing, and Internet of Things (IoT) techniques. Individuals downloaded 204 billion mobile applications in 2019 \cite{Ian_R2021}. Meanwhile, various IoT applications have been deployed \cite{Padraig_R2020} \cite{zeng_iot19} for more intelligent lifestyle, including smart home, healthcare, smart cities, etc. Mobile and IoT applications (Apps) are becoming increasingly popular globally due to their ease of use and powerful functions, which make our work and life more convenient and smarter. 

Owing to their proliferation, mobile and IoT apps also attract cybercriminals' interests \cite{Sufatrio_S2015} \cite{Balliu_J2019} \cite{zhang_tpds12}. More frequent interactions on individuals' private information occur during app's executions. According to Symantec's report in 2019 \cite{Symantec_R2019}, one in 14.5 apps accessed high-risk data without permissions from users. Attackers exploit the vulnerabilities of apps to stealthily collect sensitive user or device's information for illegal use \cite{demetriou_C2016} \cite{Grace_C2012} \cite{Surbatovich_C2017}, such as phone number, location data, credit card information, home status etc. For example, in 2020, multiple popular Android apps, such as Baidu Maps, were found leaking users' location data. Thus, data leakage has become one of the most representative security incidents in the past few years \cite{ZhouY_C2012} \cite{Rubin_C2015} \cite{McAfee_R2014} \cite{Celik_C2018} \cite{liu_comst18} \cite{nan_comst18} \cite{rory_tcyb19} \cite{miao_csur21}. And it poses a significant threat on user's identity security, financial security, or even life security.

Defensive mechanisms are proposed to protect private data from leakage including cryptography methods \cite{Chatzikonstantinou_C2016}, access control \cite{Bugiel_C2013} and information flow control \cite{Celik_C2018} \cite{Arzt_C2014} \cite{Enck_J2014}. Cryptography approaches provide a secure way to prevent the eavesdropping attack by encrypting the sensitive data, but the misuse and extra overhead limit their reliability and practicality \cite{Waye_C2017}. Access control based permission framework is an effective mechanism on limiting third-party app's access to sensitive data. It is widely used in smartphone and IoT platform \cite{Primal_C2015} \cite{Lee_C2017}. However, it has been criticized to be insufficient and undemanding for eliminating information leaks \cite{Acar_C2016}. While permissions can protect the first point of access, the user has no control over data once it is sent to another application. Unauthorized disclosure may happen through an illegal sharing data flow, which cannot be solved only by permissions. Therefore, information flow control (IFC) is proposed for long-term fine grained control over information during the whole processing procedure in an application. It is a fundamental technique in cyber security for detecting and preventing unauthorized disclosure by the enforcement of the specified flow policies \cite{Suh_C2004}. Today, there are a large number of IFC techniques applied for security checks on various apps in different platforms, from static \cite{Arzt_C2014} \cite{Li_C2015} \cite{rory_tcyb19} to dynamic \cite{Enck_J2014} \cite{Jia_C2017}, Android \cite{Wei_C2018} \cite{xiao_tifs19}, iphone\cite{Egele_C2011}\cite{Ren_C2016} to IoT \cite{Celik_C2018} \cite{Nguyen_C2019}.  In addition, unlike the Android platform, Apple iOS is not an open-source operating system which hinders the open research on the analysis and vetting of iPhone's apps. Due to the difficulties in crawling and disassembling iOS apps, few research works adapt IFC technique to address the data leakage problems in iOS platform.

Due to the increasing importance and attention on data leakage prevention in Android and IoT apps, there are already some surveys on security mechanisms adopted in mobile and IoT platforms, mainly including the Android malware detection \cite{Sufatrio_S2015} \cite{Tam_S2017}, privacy in smartphones \cite{Pennekamp_S2017} and IoT application's security \cite{Celik_S2019}. Sufatrio \textit{et al.} \cite{Sufatrio_S2015} surveyed and classified the existing works that provide security protection for Android devices. They categorized different approaches based on the apps deployment stages. Kimberly \textit{et al}. \cite{Tam_S2017} presented a survey on emerging Android malware analysis and detection techniques, along with their effectiveness against evolving malware. But information flow control was briefly introduced without systematic and in-depth analysis. Besides, recent works in the past three years were also not included in these surveys, especially for machine learning-based works. Celik \textit{et al}. \cite{Celik_S2019} studied program-analysis techniques for securing IoT apps; however, their survey only focused on IoT platform. Besides, most above surveyed works focus on the analysis of developed Apps, while IFC-based defensive mechanisms at developing stage are still missing. 

Different from the existing surveys on malware and vulnerability detection on specific platforms, this survey provides a systematic framework to incorporate different types of information flow control mechanisms for data leakage prevention in mobile and IoT apps.  We provide a new, comprehensive and insightful survey on application analysis and vetting from a specific technique view, \textit{i.e.}, IFC. Detailed descriptions and comparisons are conducted for the in-depth analysis on the features, effectiveness and performance, which are not included in the previous works.

Our survey thoroughly reviews recently published high quality research papers, which adopt different IFC techniques to prevent data leakage throughout the whole life cycle of Apps. We selected IFC-related papers from four top cybersecurity conferences, \textit{i.e.}, \textit{IEEE Symposium on  Security and Privacy  (S$\&$P)}, \textit{ACM Computer and Communications Security (CCS)}, \textit{USENIX Security Symposium}, and \textit{the Network and Distributed System Security (NDSS)}. Searching keywords were \textit{information flow},  \textit{data leakage}, \textit{Android} and \textit{IoT}. We also studied the references in the selected paper and the papers refering to them. Through this investigation, we aim to give a comprehensive and in-depth introduction on our topic. Therefore, a systematic IFC technique framework, \textit{i.e.},  IFC based defensive chain, is established for lifecycle sensitive data protection in mobile and IoT apps covering all stages including design, development, deployment, execution and maintenance. 

The contributions of our survey are listed as follows:

Firstly, we collect and study more than 150 papers published in recent three years for the investigation on the research pattern and trend of IFC techniques used in mobile and IoT applications. There are few works in iOS apps, so we mainly focus on works about Android and IoT platform in this paper.

Secondly, the principal contribution is that we design a systematic IFC technique framework for mobile and IoT apps based on the collection and investigation of state-of-the-art schemes, methods and experiments. In this framework, the surveyed works are divided into four categories according to the development stages in the life cycle of an app, \textit{i.e.}, model time, develop time , deploy time and run-time IFC. It can provide a clear and thorough summary on current works for an easy understanding to the new researchers in this field. The systematic technique framework for integrating different IFC approaches is summarized in Section \ref{sec_method}. 

Thirdly, the IFC techniques adopted by each work are detailed and compared in the form of tables and figures. All the related works are categorized into the above four phases. Besides, for deploy-time and run-time IFC, they are further divided into 3 to 4 sub-classes. For similar works in each category/class, we first briefly introduce some basic works in the field and describe core works published in recent years in detail, including their research goals, main processing procedures and evaluations. We also make in-depth comparisons among different works, including their main techniques, features, performances, and so on. Through the thorough review, we can provide a clear description of recent developments of IFC used for detecting data leakage in mobile and IoT apps.

Finally, challenges and future directions are also discussed in the end. We summarize the challenges for the different types of techniques in each category. Based on these challenges, we aim to provide valuable insights to the research community for further research.

This survey is organized as follows. Section \ref{sec_method} summarizes the overview of the lifecycle oriented IFC technique framework, including the construction of a defensive chain and research methodology. Section \ref{sec_LR} details existing works in the field of IFC for data leakage detection in mobile and IoT applications. Based on the above description, Section \ref{sec_discussion} discusses the challenges and future direction in this area. Finally, Section \ref{sec_conclusion} concludes the paper.

\section{Information Flow Based Defensive Chain for Data Leakage Prevention in Moible and IoT Applications} \label{sec_method}
Before constructing the technique framework, we first analyze the security challenges on mobile and IoT applications and focus on the data leakage threats. Then based on the pattern of software development and management, i.e., application's development cycle, we construct an IFC defensive chain, which integrates main types of IFC techniques in different phases for comprehensive protection on users' private data.  

\subsection{Security Challenges on Android and IoT Applications}
Compared with traditional applications in personal computers (PCs), mobile and IoT applications have become the main attack targets due to their prevalence in current world. Android or IoT apps usually run on a customized operating system or platform, which provides different application lifecycle management modes. 

\begin{figure}[t]
\setlength{\abovecaptionskip}{0.2cm}
\centerline{\includegraphics[scale=0.5]{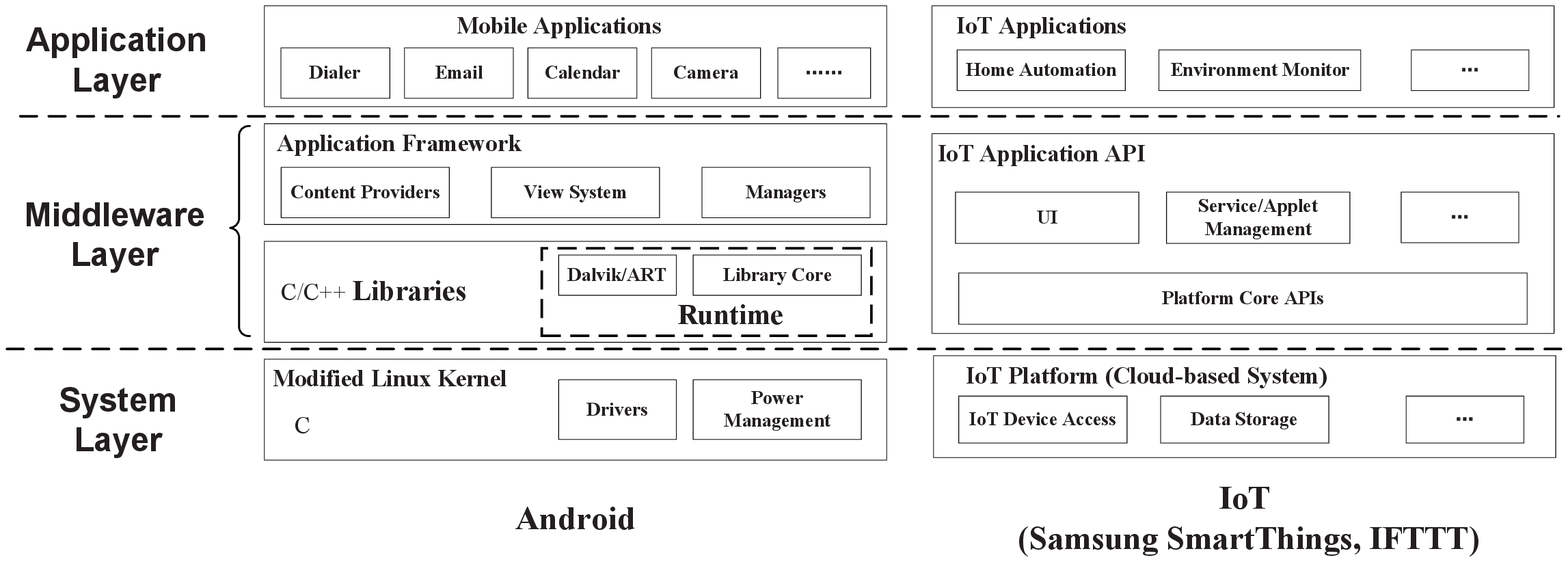}}
\caption{The General Architecture of Mobile and IoT Applications}
\label{fig_app_structure}
\vspace{-0.2cm}
\end{figure}

\textbf{Android Apps}. Android apps are written in Java. Regular Java programs in PC usually have a single entry point, \textit{i.e.}, \textit{main} function in program. On the other hand, Android apps can have multiple entry points for quick switch among different apps. Then, one app's code is compiled into Dalvik bytecode (.dex) for execution. Finally, the bytecode file is packaged with related resources and data files into an APK (Android Package) file for installation. For efficient management on apps in a platform with limited resources, an Android app is composed by four different types of basic components, \textit{i.e.}, \textit {activity}, \textit{service}, \textit{broadcast receiver}, and \textit{content provider}. Android developers write the code with reference to the lifecycle methods of app components. An application process's lifetime is not directly controlled by the application itself. The Android framework interacts with various application components independently and calls one component’s lifecycle methods to execute the app based on \textit{Davilk virtual machine} or \textit{Android Runtime}. The architecture of Android applications is shown in Fig.\ref{fig_app_structure}. 

\textbf{IoT Apps}. IoT apps are usually deployed on a cloud-based or local IoT platform, such as Apple's HomeKit \cite{Apple_HomeKit}, OpenHAB \cite{Open_Hab}, Samsung’s SmartThings \cite{Smart_things}, and Android Things \cite{Android_Things}. These apps can connect devices in a physical environment through an edge gateway, use cloud backend to compute and remotely control devices. Therefore, data collection, control, and interoperability are three crucial functions in these applications. There are diverse developing languages due to a variety of application fields and platforms, e.g., groovy language for SmartThings, and Java for Android Things. Further, some apps are written in a Domain Specific Language (DSL). Besides, some IoT apps may run in a secure and efficient sandbox. The architecture of IoT applications is also shown in Fig.\ref{fig_app_structure}.

Although mobile and IoT apps may be developed in different ways, they have similar abstract application structures. These applications are usually built on an underlying system/platform which has access to different resources such as device, data, and network. There is a middleware layer between the application and system/platform. The middleware layer constructs the application framework. It provides various API to access the resources in the system/platform, which can mask the complex operations on system/platform. For applications at the top layer, they complete the service function according to users' requirements. For a clear description, we call mobile and IoT apps as \textbf{smart applications} to distinguish traditional computer applications in this paper.

Smart applications are more prone to data leakage due to insufficient security protection mechanisms. Various attacks are targeting on different levels. For application-level attacks, they can eavesdrop users' private data through the vulnerabilities in the application codes. For middleware-level and system-level attacks, they try to access users' data through the application framework or the system, such as illegal use on authorised APIs, the attempt on gaining extra privileges. Furthermore, the complex business process of applications, multi-level development and execution environments make it becoming more difficult for developers, market managers or customers to detect the vulnerabilities and malicious behaviours in the application. 

\subsection{A lifecycle Information Flow Defensive Chain on Smart Applications}
\begin{figure}[t]
\setlength{\abovecaptionskip}{0.2cm}
\centerline{\includegraphics[scale=.3]{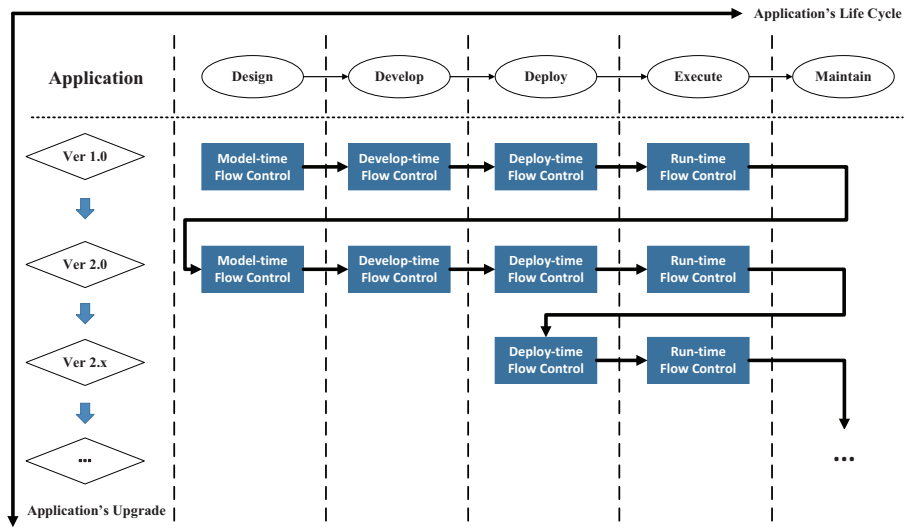}}
\caption{Information Flow Defensive Chain over Application Life Cycle}
\label{fig_defensive_chain}
\vspace{-0.4cm}
\end{figure}

As we all know, the life cycle is the basic concept for describing standard software development and management. For smart apps, there are five main phases during their development and evolution, i.e., design, development, deployment, execution and maintenance. Although data leakage only occurs during the execution of an application, the vulnerability has existed since the design phase. As our survey shows, each IFC approach used in the specific phase has its advantages and disadvantages. A single security mechanism in one specific phase can not guarantee the prevention of data leakage over the whole application's life cycle. By resolving the application life cycle into a two-dimension software evolution process, we construct a continuous chain-style IFC technique framework for prevention on users' data leakage in smart applications, i.e., lifecycle Information Flow Defensive Chain. Fig.\ref{fig_defensive_chain} shows the defensive chain.

lifecycle information flow defensive chain is a security protection strategy that addresses the threat of data leakage by combing different IFC techniques in different phases. It is a continuous process, checking application's model, parsing and analyzing the application code for early detection on data leakage, testing and monitoring the execution of the application for real-time and precise protection, and restarting the defense process when the application is updated. The chain-style defensive framework aims to provide continuous protection along smart application's development and evolution procedure. According to security techniques in different phases, the information flow defensive chain consists of four main defensive clusters, i.e., \textbf{model-time flow control, develop-time flow control, deploy-time flow control and run-time flow control}. 

\subsubsection{\textbf{Model-time flow control in the design phase}}\label{sub_sub_sec_model}
Model-time flow control is the first phase of the defensive chain. In software engineering, developers first use the modeling techniques to formally represent methods, functions, objects and the overall structure and interactions so that the resulting functionality satisfies users' requirements\cite{fowler_B2004}. During the modeling process, developers use abstract language and diagrams for a clear description and ease of understanding of users' requirements. However, inappropriate design of the application, such as the class's private member, insecure access interface, or wrong message flow, may leave the vulnerabilities in the following application development.

To eliminate the leakage vulnerabilities at an early stage, developers can integrate the information flow (IF) properties into the modeling procedure\cite{alghathbar_J2006} \cite{seehusen_J2009}. By using standard modeling language and diagrams, IF-related properties on user's data are also introduced into different classes' views as the nonfunctional properties. Then we can restrict the information flows within the application model by IFC techniques. We can also check whether the application's behaviour (including method, action, etc.) violates the security requirements with the automatic verification approaches, e.g., model checking \cite{clarke_B2018} \cite{barthe_J2011}, automata\cite{Dsouza_J2005}, etc..

\subsubsection{\textbf{Develop-time flow control in the development phase}}\label{sub_sub_sec_Compile}
Programmers design and develop the application based on users' requirements. The application's vulnerability is one of the primary cause of data leakage. Therefor, application developers want to secure their apps before official release. They can adopt the information flow analysis technique for early detection of the data leakage caused by the complex program logic, developers' careless, and so on. 

During the development, security checking based on type system is directly running on the application's source code, which is widely used for information flow control. A type system denotes a set of rules in programming languages, . The rules assign attributes called types to various components in a program, including variables, expressions, functions or modules \cite{volpano_J1996}. Type-based security check mechanisms introduce security labels that abstract program values as types based on security lattice model, track dependencies within a program and prohibit sensitive data leakage through public outputs \cite{Rajani_J2017}. In these mechanisms, researches extend the original programming language to support the secure-flow type system, and build them into the compilers or interpreters for develop-time or run-time checking. Through its develop-time analysis, they can detect insecure flows and prevent data leakage at an early stage. 
 
\subsubsection{\textbf{Deploy-time flow control in the Deployment Phase}}\label{sub_sub_sec_deploy}
After the development, the application is released on app markets for different systems/platforms, such as Google Store, Apple Store and SmartThings store. And these apps can be downloaded and deployed on user's device. Few developers may execute the verification procedure due to the high development cost. Besides, some developers may be malicious attackers themselves. Therefore, defenses after the development phase play a more essential and vital role. 

During app's deployment, static analysis is widely used for information flow security verification without executing the application. It performs taint or data dependency analysis on  source or compiled code, to identify the leakage path of sensitive data, such as FlowDroid \cite{Arzt_C2014}, DroidSafe \cite{Gordon_C2015}, SAINT \cite{Celik_C2018}. The taint analysis statically derives a set of possible data which is tagged with a tainted label, and decides whether sensitive data transfer to untrusted channels according to the propagation rules based on static analysis on applications' code. The data dependency analysis constructs the dependency graph of different variables in the application, and translates the information flow security into a reachability problem. For static analysis, it can achieve a high code coverage via analyzing all resources or possible paths of the application. However, it faces critical challenges including handling code obfuscation, dynamically loading code, reflection calls, native code, and multithreading issues without actual execution paths. Thus, rum-time analysis techniques are also an essential part of IFC defenses.

\subsubsection{\textbf{Run-time flow control in the Execution Phase}}\label{sub_sub_sec_execution}
In this phase, the application runs on a system/platform to complete the user's tasks. Attackers can leverage the dynamic malicious codes and vulnerabilities to avoid static flow analysis. So run-time analysis and monitoring mechanisms are also needed for enforcing a more robust policy to detect data leakage, which can check the illegal behaviours dynamically and can overcome the drawbacks of the static flow analysis.

Dynamic analysis is in contrast to static analysis, which executes a program and observes the run-time behavior. It provides limited code coverage, because only one path is shown per execution, which, however, can be improved with stimulation. The dynamic taint analysis technique is one of the most common techniques for tracking the information flows within applications at run-time. The dynamic taint analysis labels (taints) sensitive data from specific sources and handle labelled transitions (taint propagation) between variables, files, and procedures at run-time. If a tainted data transmits out of smart devices/platforms via some functions (sinks), then we can monitor the data leakage dynamically. Besides, some researchers propose secure multi-execution mechanism(SME) to overcome the limited coverage and improve the precision on detection of data leakage during execution time. Instead of preventing insecure flows in the application, SME aims at 'repairing' them on the fly in contrast to the monitoring and tracking techniques. This approach achieves security by separation of executions at different security levels, which means executing an application multiple times based on specific rules for I/O operations \cite{Rafnsson_S2016}. Static and dynamic analysis have their own advantages and disadvantages. Some researchers propose hybrid approaches by combining Deploy-time and Run-time defensive mechanisms, i.e. static and dynamic analysis, to increase robustness, precision and efficiency. 

\subsubsection{\textbf{Flow Control in the maintenance phase}}\label{sub_sub_sec_maintenance}
For benign applications, code is updated for function upgrade or error repair from time to time. Some vulnerabilities can be eliminated during the maintenance while new functions may also introduce new vulnerabilities. For malicious applications, the malicious code is updated for better concealing to avoid being detected.

There are two different defense patterns for updated apps included in the IFC defensive chain. On the one hand, if the app is updated to a new version, the app may be redesigned and redeveloped. That means defensive mechanisms should be restarted from the design phase, and all the above IFC techniques could be adopted for leakage detection in new-version applications. On the other hand, model-time and develop-time defensive mechanisms sometimes may be ignored due to the extra time and development cost, especially for those applications with minor changes. Besides, malicious developers will bypass these mechanisms anyway. Therefore, it is essential for users to adopt the deploy-time and run-time security mechanisms to detect new vulnerabilities or malicious codes.

\section{IFC-based Defensive Mechanisms on Smart Applications} \label{sec_LR}
\begin{figure}[t]
\setlength{\abovecaptionskip}{0.2cm}
\centerline{\includegraphics[scale=.5]{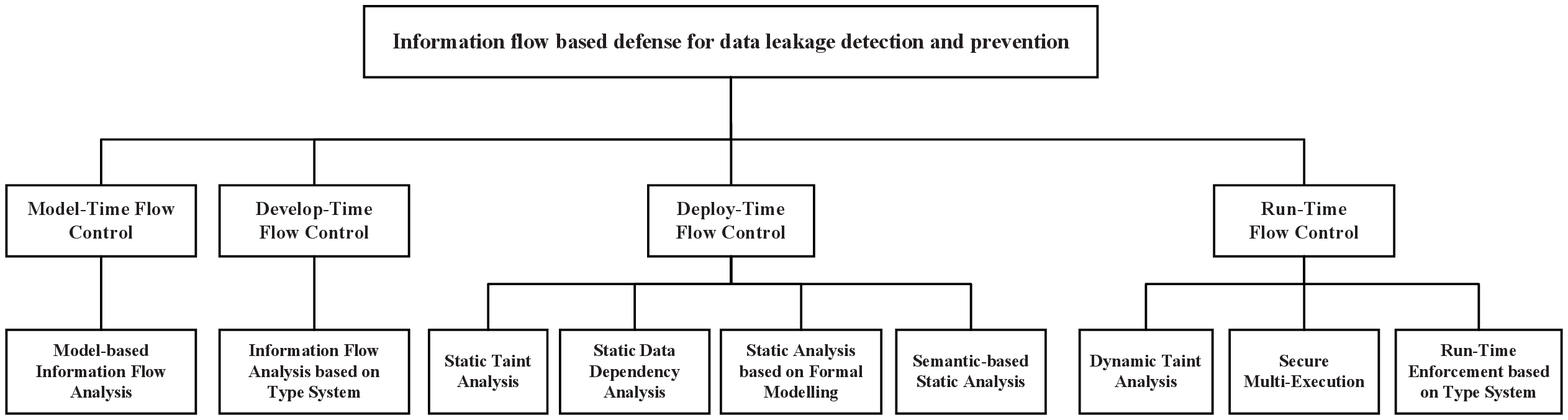}}
\caption{Categories of Information Flow Based Analysis Approaches}
\label{fig_categories}
\vspace{-0.4cm}
\end{figure}

This section details various IFC techniques in different phases. We group them into four categories, and plot our taxonomy tree, as shown in Fig. \ref{fig_categories}. 

We further classify these works based on the employed techniques, and constitute the second level of the tree. Works in category 3 and 4 constitute the majority of the surveyed works (22 of 26), indicating two hot research topics in IFC, namely deploy-time analysis and run-time analysis approaches. Table \ref{table_approache_phases} lists the investigated solutions and describes their main characteristics. Most work (19 of 26) focus on leakage in Android due to its widespread usage. Vulnerabilities and malware are two main security threats. There are various techniques adopted for different phases. 1) In the Design phase, model-based analysis is studied for early elimination on leakage vulnerabilities in the app's formal model. 2) Type system can be adopted to check the app source code during the Develop phase. 3) In the Deploy phase, static-analysis based approaches are proposed for developed apps. Bytecode, Apk files and source code are three main types of analysis objects. 4) Run-time checking mechanisms provide precise and instant ways to detect data leakage. Execution instructions and source code are two main types of analysis objects. Furthermore, most works focus on explicit and implicit flows. Few works also consider termination and timing flows, which partly support resistance to side channel attack. Main contributions in each work are detailed in accord with the taxonomy.

\begin{table}[t]
\setlength{\abovecaptionskip}{0.2cm}
\footnotesize
\caption{IFC-based Defensive Approaches in Different Phases}
\centering
\resizebox{\textwidth}{55mm}{\begin{tabular}{cccm{3.5cm}m{3cm}m{3.5cm}m{3cm}}
\toprule
\textbf{Work} & \textbf{Platforms} & \textbf{Attack Models} & \makebox[3.5cm]{\textbf{Main Techniques}} & \makebox[3cm]{\textbf{Analysis Objects}} & \makebox[3.5cm]{\textbf{Flow Types}} & \makebox[3cm]{\textbf{Used Tools}}\\
\midrule

\multicolumn{7}{c}{\textit{\textbf{1. Model-Time Flow Control in Design Phase}}}\\[3pt] 
UML-IF Modeling\cite{Katkalov_C2015}   & Android & Vulnerabilities & UML modeling & UML Model & Explicit Flow, Implicit Flow & -\\[3pt] 
CPS-IF Modeling\cite{Geismann_C2018}   & IoT & Vulnerabilities & Domain-specific modeling & Domain-specific Model & Explicit Flow & -\\[3pt] 
\midrule

\multicolumn{7}{c}{\textit{\textbf{2. Develop-Time Flow Control in Develop Phase}}}\\[3pt] 
Android IFC\cite{Chen_C2018}  & Andorid & Vulnerabilities & Type System & Java Code & Explicit Flow, Implicit Flow & - \\[3pt]
IoT IFC\cite{Prasad_C2020} & IoT    & Vulnerabilities & Type System & Domain-specific Code & Explicit Flow, Implicit Flow & - \\[3pt]
\midrule

\multicolumn{7}{c}{\textit{\textbf{3. Deploy-Time Flow Control in Deploy Phase}}}\\[3pt] 
IccTA\cite{Li_C2015}  & Andorid & Vulnerabilities, Malware & Static Taint Analysis & Dalvik bytecode &  Explicit Flow, Implicit Flow & Dexpler, IC3, FlowDroid \\[3pt]

JN-SAF\cite{Wei_C2018} & Andorid & Vulnerabilities, Malware & Static Taint Analysis & Dalvik bytecode and Native binary code &  Explicit Flow, Implicit Flow & AmAndroid, angr, jpy\\[3pt]

SAINT\cite{Celik_C2018} & IoT & Vulnerabilities, Malware & Static Taint Analysis & Groovy Code &  Explicit Flow, Implicit Flow (partly) & Groovy compiler, Groovy Swing console\\[3pt] 

DroidSafe\cite{Gordon_C2015}  & Andorid & Vulnerabilities, Malware & Data Dependency Analysis & Dalvik bytecode or Source Code, Android environment &  Explicit Flow & Soot \\[3pt]

R-Droid\cite{Backes_C2016}  & Andorid & Vulnerabilities, Malware & Program Slicing & Dalvik bytecode &  Explicit Flow, Implicit Flow & JOANA\\[3pt]

HornDroid\cite{Calzavara_C2016}  & Android & Vulnerabilities, Malware &Horn clauses, Static Taint Analysis  & Dalvik bytecode &  Explicit Flow & Z3 (SMT solvers)\\[3pt]

SUPOR\cite{Huang_C2015}  & Android & Vulnerabilities & Semantic Analysis, NLP, Static Taint Analysis& Apk & - & Apktool, dexlib and WALA, Eclipse plugins, Dalysis\\[3pt]

FlowCog\cite{Pan_C2018}  & Android & Vulnerabilities, Malware & Static Taint Analysis, Semantic analysis, Dynamic Analysis (Optional), NLP  & Apk & Explicit Flow, Implicit Flow & FlowDroid, Soot, Stanford Parser \\[3pt]

ClueFinder\cite{Nan_C2018}  & Android & Vulnerabilities, Malware & NLP, SVM, Static Taint Analysis & Dalvik bytecode & Explicit Flow & FlowDroid, Stanford Parser\\[3pt]

AAPL\cite{Lu_C2015}  & Android & Vulnerabilities, Malware & Static Taint Analysis, Peer Voting & Dex bytecode & Explicit Flow, Implicit Flow & Dalysis and IBM WALA\\[3pt]

MUDFLOW\cite{Vitalii_C2015}  & Android & Malware & Static Taint Analysis, SVM & Apk & Explicit Flow & FlowDroid, SVM\\[3pt] 

\midrule
\multicolumn{7}{c}{\textit{\textbf{4. Run-Time Flow Control in Execution Phase}}}\\[3pt] 
TaintART\cite{Sun_C2016}  & Android & Vulnerabilities, Malware & Dynamic Taint Analysis & Dalvik Instruction & Explicit Flow & dex2oat \\[3pt]

TaintMan\cite{You_J2017}  & Android & Malware & Static Instrumentation, Reference Hijacking, Dynamic Taint Analysis & Apk, Dalvik Instruction & Explicit Flow, Implicit flow, Terminiation Flow & Apktool, Smali/Baksmali \\[3pt]

PIFT\cite{Yoon_C2016}  & Android & Malware & Dynamic Taint Analysis, Predictive Heuristics & ARM Instruction & Explicit Flow & - \\[3pt] 

CDroid\cite{Wu_J2018}  & Android & Vulnerabilities, Malware & Formal Model, Dynamic Taint Analysis & Dalvik Instruction & Explicit Flow & TaintDroid \\[3pt]

NDroid\cite{Xue_J2019} & Android & Vulnerabilities, Malware & Dynamic Taint Analysis, Static Taint Analysis & Dalvik Instruction, ARM /Thumb /Thumb2 Instruction & Explicit Flow & QMENU, TaintDroid, darm\\[3pt]

ContexIoT\cite{Jia_C2017}  & IoT & Vulnerabilities, Malware & Code Patching, Static Analysis, Dynamic Taint Analysis & Groovy Source Code, Instruction & Explicit Flow, Implicit Flow & - \\[3pt]
 
FlowFence\cite{Fernandes_C2016}  & IoT & Malware & Opacified Computation, Dynamic Taint Analysis & Java Source Code & Explicit Flow, Implicit Flow, Termination Flow & - \\[3pt]

Weir\cite{Nadkarni_C2016}  & Android & Malware & Lazy Polyinstantiation & - & Explicit Flow, Implicit Flow, Terminiation Flow & overlayFS, Linux Security Module \\[3pt]

Ariel\cite{Chakraborty_C2019}  & Android & Malware & Secure Multi Exeuction, Sandbox & - & Explicit Flow, Implicit Flow & - \\[3pt]

FlowIT\cite{Bastys_C2018}  & IoT & Vulnerabilities, Malware & Runtime Type Checking & JavaScript Code & Explicit Flow, Implicit Flow, Terminiation Flow, Timing Flow & JSFlow \\[3pt]

CaITApp\cite{Balliu_C2019}  & IoT & Vulnerabilities, Malware & Runtime Type Checking & IoT apps' model & Explicit Flow, Implicit Flow & - \\[3pt]

\bottomrule
\end{tabular}}
\label{table_approache_phases}
\vspace{-0.4cm}
\end{table}

\subsection{Model-time and Develop-time information flow control}
Mobile and IoT apps become the targets of a number of security threats. Information flow vulnerabilities have been introduced from the design and coding phase. This type of vulnerabilities is more complicated and difficult to detect after the development phase. Therefore, Model-time and Develop-time IFC techniques provide a promising way to eliminate these vulnerabilities at an early stage. If the possible vulnerabilities can be addressed and fixed during app's early development phase, the potential attack can be minimized. But these approaches may be skipped by the malicious apps which are intended to steal user's private information.    

\subsubsection{\textbf{Model-based information flow control.}}
Software modeling is widely used in the software design phase for precisely describing the software's constitutions, functions, workflows, \textit{etc}. \cite{Ramos_J2012}. For different modeling languages and platforms, there are different model-based security approaches integrated into the modeling process, such as UMLsec \cite{Jan_C2002}, Model-driven Security \cite{Basin_J2006}, SysML-Sec\cite{Roudier_C2015}, which support formal analysis and verification on software's security properties. Information flow security is one of the most critical security properties supported by these approaches\cite{Tuma_C2019}. Considering the features of Android and IoT applications, some recent model-based information flow works are studied further \cite{Katkalov_C2015}, \cite{Geismann_C2018}.

\textit{a. Model-based IFC on Android apps.} For Android apps, large-scale interactions between mobile apps and web services are very common, which makes protecting sensitive information is a difficult task. Katkalov \textit{et al.} \cite{Katkalov_C2015} proposed a model-driven approach to develop distributed applications with steady information flow (IF) guaranteed. They proposed to express IF properties by using a graphical UML notation. Multiple IF properties on the model were automatically verified by model-based interactive verification and static analysis approaches. 

In Katkalov' approach, the UML model of the application was firstly created according to the functional requirements. Then the IF properties were introduced into the model by using UML activity diagrams. Declassification, trace and platform-specific properties were also defined for flexible, fine-grained information flow protection. After that, they generated a code skeleton and a formal model automatically for the verification of different IF property. The transitive noninterference was checked on the code level automatically through static analysis methods. Other IF properties, including the intransitive noninterference, declassification and trace, were guaranteed by the verification on the formal model. Discussions of the formal model, framework and verification techniques in their approach are detailed in \cite{Stenzel_J2014}. For evaluations, they showed a travel planner case study as an example modeling with IF properties guaranteed.

\textit{b. Model-based IFC on IoT apps.} Different from PC and Android apps, IoT apps are developed on smart embedded devices which contain lots of interactions with the physical environment or humans. Cyber-physical systems (CPS) is a typical IoT application. Geismann \textit{et al.} \cite{Geismann_C2018} proposed a novel secure-by-design process for CPS based on model systems engineering. By specifying the security requirements and integrating dedicated threat models, engineers can ensure security of the systems under development by design.

Their design process mainly consists of three steps, \textit{i.e.}, interdisciplinary system design, platform-independent software engineering and platform-specific software engineering. Firstly, an interdisciplinary system was modelled based on the CONSES (CONceptual design Specification technique for the Engineering of complex Systems) technique. It provided a set of security-relevant views on a system while in developing \cite{Anacker_J2014}. IF property was included in the system model. Then a software component model was derived from the system model. During the derivation, the information flow policy was decomposed into local flow policies for restricting components. Based on the model and security policies, each component was verified by IFC techniques \cite{hedin_J2012} against the local flow policies. When a violation occurs, the software engineer can track the error by adjusting the flow strategy or correcting the messaging behavior. 

For clear description, they used an autonomous car model integrating information flow control policies as an example to illustrate the secure-by-design process for CPS. Although the Model-time information flow analysis approaches can detect and eliminate early security flaws in software, the effectiveness and precision are limited by the modeling process. They are usually applied as auxiliary mechanisms and less studied than those in other phases.

\subsubsection{\textbf{Information flow analysis based on type system.}} 
Information flow analysis built on type system is the most widely used approach for checking code's security properties at the compile time \cite{volpano_J1996} \cite{Sabelfeld_J2003}. Researches have proposed several type systems for ensuring information flow security in Android and IoT applications \cite{Chaudhuri_C2009}, \cite{Ernst_C2014} \cite{Chen_C2018} \cite{Khan_J2019} \cite{Prasad_C2020}.  \cite{Chaudhuri_C2009} firstly presents a core typed language to ensure data-flow security in Android applications  Some type-based analysis techniques are also applied by the app store owner\cite{Ernst_C2014}. Besides, more frequent interactions among different components (within an app or among multiple apps) become a major cause on leakage compared with traditional PC programs. Here we study two recent works, i.e., \cite{Chen_C2018} and \cite{Prasad_C2020}, to illustrate the precise IFC across multiple components in Android and IoT applications.

\textit{a. Type system for IFC in Android}. In Android, if designing an information flow type system, conventional type systems usually take all branches as possible leakage for conservative protection. Therefore, Chen \textit{et al.} \cite{Chen_C2018} designed a more precise type system under the Android framework's access control model. Their type system was based on Banerjee and Naumann's \cite{Banerjee_J2005} work (BN system).  

In their type system, an application was modeled as a group of functions with a statically assigned permission set. For permissions, they focused on the inter-component communications within and across apps. Then they extracted the operational semantics of expressions and commands. The semantics improved existing work \cite{volpano_J1996} by extending functions and an operator for permission checks. After that, security types were defined as mappings from permissions to security levels in a partial order. Next, a series of typing rules for expressions, commands and functions were given for security check on the program. Most of them were similar to traditional information-flow type systems \cite{volpano_J1996}, \cite{Sabelfeld_J2003} and \cite{Banerjee_J2005} except for those on branches with permissions and function calls. In addition, they introduced a refined merging operation to reduce the false positives. Finally, they proved the soundness of the designed type system\cite{Chen_C2018}. 

\textit{b. Type system for IFC in IoT.} IoT apps face the security threats from insecure definition of interfaces and overlook of security classification of data. \cite{Prasad_C2020} proposed a security type system which integrates Denning's lattice-based secure information flow (SIF) framework into LUSTRE, which is a high-level abstract programming model for IoT apps. \\ 
\indent In the proposed type system, they normalized the core syntax of LUSTRE, formulated appropriate rules for LUSTRE equations, and represented the security types and constraints as a symbolic style. They also gave a definition of security for LUSTRE programs based  on the security typing rules. They established the soundness of the type system based on Volpano's approach \cite{volpano_J1996}. Technical analysis results proved that their security type system is robust with LUSTRE.

\subsection{Deploy-time Information Flow Control}
Model-time and Develop-time control are not widely used because of the inconsistency between model and program codes, the vulnerabilities during coding, the development cost and so on. Besides, malicious application developers will try to evade these mechanisms. Therefore, after-developement IFC mechanisms are essential and widely studied. For developed apps, static analysis is widely used to the detect insecure information flows without app's execution. The analysis object is usually the intermediate representation (IR) of the program generated by the source or compiled code. 

\begin{figure}[t]
\setlength{\abovecaptionskip}{0.2cm}
\centerline{\includegraphics[scale=.4]{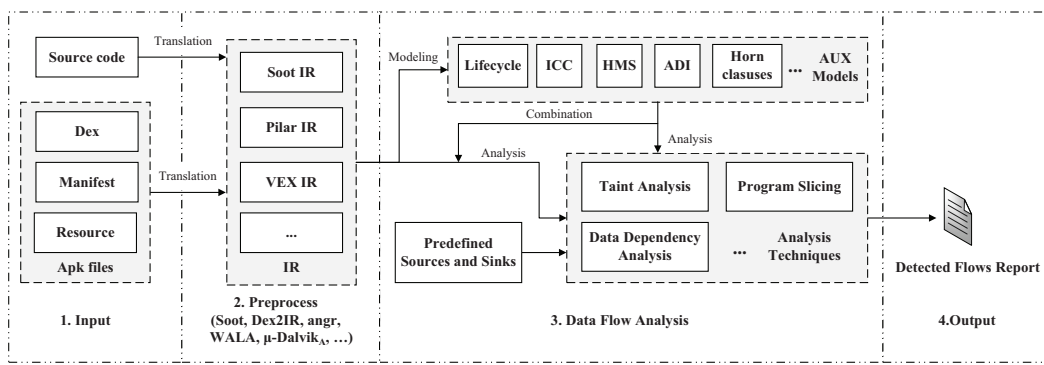}}
\caption{Static Information Flow Analysis}
\label{fig_static_IF_analysis}
\vspace{-0.4cm}
\end{figure}

The general static analysis procedure is shown in Fig.\ref{fig_static_IF_analysis}. Source code or apk files are first translated into different forms of IRs. Then data flow analysis techniques are performed on these IRs. For improving the accuracy of the analysis, some works also build auxiliary models to describe the features or behaviours between the applications and system. Based on predefined sources and sinks, detected flows are reported to users. 

The static information flow analysis research works in recent years on smart applications can be categorized into two main groups according to different research goals. One is to \textit{improve the static analysis techniques for precisely capturing the flows across complex app's code}. The other is to \textit{improve the validation techniques for more accurate decision  on the legality of the captured flows}.

We detail recent six representative works which aim at improving the static analysis technique in our survey \cite{Li_C2015} \cite{Wei_C2018} \cite{Celik_C2018} \cite{Gordon_C2015} \cite{Backes_C2016} \cite{Calzavara_C2016}. These works are summarized in Table \ref{table_static_analysis}. For different works, we compare them with seven criteria including \textit{ main technique}, \textit{application's type}, \textit{developing language}, \textit{IR}, \textit{analysis sensitivity}, \textit{features} and \textit{evaluation}. We use \textit{sensitivity} to express the grain of the tracking techniques, which mainly includes \textit{field-}, \textit{object-}, \textit{context-}, \textit{flow-} and \textit{path-sensitive} techniques. And different analysis features are also considered in our comparison, \textit{i.e.}, \textit{ICC}, \textit{native code}, \textit{implicit flow}, \textit{reflection}, \textit{concurrency} and \textit{soundness}. Besides, we also compare these works in experiment environment, data set, effectiveness, and performance.  In our survey, we use their evaluation results about \textbf{TP} (True Positive), \textbf{TN} (True Negative), \textbf{FP} (False Positive), and \textbf{FN} (False Negative) to evaluate the effectiveness of different works. For above 6 works, we calculate their \textit{precision} and \textit{recall}. The average analysis time on each application is also recorded in Table \ref{table_static_analysis} to show their performance.

\begin{table}[t]
\setlength{\abovecaptionskip}{0.2cm}
\footnotesize
\newcommand{\tabincell}[2]{\begin{tabular}{@{}#1@{}}#2\end{tabular}}
\centering
\caption{Static Information Flow Analysis Approaches}
\resizebox{\textwidth}{50mm}{
\begin{tabular}{m{0.1cm}lm{2cm}<{\centering}m{2cm}<{\centering}m{2cm}<{\centering}m{2cm}<{\centering} m{2cm}<{\centering} m{2.5cm}<{\centering}}
\toprule
\multicolumn{2}{c}{Work} & \textbf{IccTA} \cite{Li_C2015} & \textbf{JN-SAF} \cite{Wei_C2018} & \textbf{SAINT} \cite{Celik_C2018} & \textbf{DroidSafe} \cite{Gordon_C2015} & \textbf{R-Droid} \cite{Backes_C2016} &  \textbf{HornDroid} \cite{Calzavara_C2016} \\
\midrule
\multicolumn{2}{c}{\textbf{Technique}} & Taint Analysis &  Taint Analysis & Taint Analysis & Data Dependency & Program Slicing & \tabincell{c}{Formal modeling \\ $\&$ Taint Analysis} \\
\midrule 
\multicolumn{2}{c}{\textbf{App Type}}  & Android App & Android App & IoT App & Android App & Android App & Android App\\
\midrule
\multicolumn{2}{c}{\textbf{App Language}}  & Java & \tabincell{c}{Java, \\ C/C++(Native Code)} & Groovy & Java & Java & Java \\
\midrule
\multicolumn{2}{c}{\textbf{App's IR}}  & Soot IR & Soot IR, HMS$^2$ & SAINT IR & Soot IR, ADI & WALA IR, light-weight ADI & $\mu$-Dalvik$_A$ Model, Horn clauses \\
\midrule
\multirow {5}{*}{\rotatebox{90}{\textbf{Sensitivity}}} 
 & Field    & $\checkmark$  & $\checkmark$  & $\checkmark$  & $\checkmark$  & $\checkmark$  & $\checkmark$ \\
 & Object   & $\checkmark$  & $\checkmark$  & $\checkmark$  & $\checkmark$  & $\checkmark$  & $\checkmark$ \\
 & Context  & $\checkmark$  & $\checkmark$  & $\checkmark$  & $\checkmark$  & $\checkmark$  & $\checkmark$ \\
 & Flow     & $\checkmark$  & $\checkmark$  & $\checkmark$  & $\times$      & $\checkmark$  & $\bullet$ \\
 & Path     & $\times$      & $\times$      & $\checkmark$  & $\times$      & $\checkmark$  & $\checkmark$ \\
\midrule
\multirow {7}{*}{\rotatebox{90}{\textbf{Features}}} 
 & ICC$^1$       & $\checkmark$ & $\checkmark$  & -         & $\checkmark$  & $\times$      & $\checkmark$ \\
 & Native Code   & $\times$     & $\checkmark$  & -         & $\times$      & $\times$      & $\times$ \\
 & Implicit Flow & $\checkmark$ & $\checkmark$  & $\bullet$ & $\times$      & $\checkmark$  & $\times$ \\
 & Reflection    & $\bullet$    & $\times$      & $\bullet$ & $\bullet$     & $\times$      & $\bullet$  \\
 & Concurrency   & $\bullet$    & $\bullet$     & -         & $\bullet$     & $\checkmark$  & $\bullet$\\
 & Soundness     & $\times$     & $\times$      & $\times$  & $\times$      & $\times$      & $\checkmark$\\
\midrule
\multirow {12}{*}{\rotatebox{90}{\textbf{Evaluation}}}
 & Environment   & Computer, JVM & Server & Computer & Server, JVM & Server & Server \\
\specialrule{0em}{2pt}{1pt}
 & Capability    & i7 CPU, 8GB heap size & 48 Cores, 256GB RAM & 2 Cores, 8GB RAM & E5-2690v2 CPU, 64GB heap size & 8 Cores, 768GB RAM & 64 Cores, 758GB Memory \\
\specialrule{0em}{2pt}{1pt}
 & Benchmark Set & DroidBench (22), ICC Bench (9) & Bench apps (22) & - & DroidBench (94) & DroidBench & DroidBench (120)\\
\specialrule{0em}{2pt}{1pt}
 & Precision$^3$ & $96.6\%$ & $90.5\%$ & - & $87.6\%$ & $97\%$ & $94\%$ \\
\specialrule{0em}{2pt}{1pt}
 & Recall$^4$ & $96.6\%$  & $100\%$ & - & $93.9\%$ & $100\%$ & $96\%$ \\
\specialrule{0em}{2pt}{1pt}
 & Real-World Apps & \tabincell{c}{GooglePlay \\ (18)} & AndroZoo (10000) & SmartThings (230) & DARPA APAC (24) & GooglePlay (22700) & \tabincell{c}{GooglePlay \\ (2 lagre ones)}\\
\specialrule{0em}{2pt}{1pt}
 & Performance (s/app)$^5$ & 25 & 88.98 & 23$\pm$5 & [261,1477] & 1560 & 2700 \\
\bottomrule\\
\multicolumn{8}{c}{$\checkmark$ = comprehensive/sophisticated support, $\bullet$ = basic support, $\times$ = not support, -= not disscussed}\\
\multicolumn{8}{c}{$1$: ICC = Inter Component Communication, $2$: HMS = Heap Manipulation Summary}\\
\multicolumn{8}{c}{$3$: Precision = TP/(TP+FP), $4$: Recall = TP/(TP+FN), TP: True Positive, FP: False Positive, FN: False Negative}\\
\multicolumn{8}{c}{$5$: s/app: average analysis time(second) on each application}
\end{tabular}}
\vspace{-0.4cm}
\label{table_static_analysis}
\end{table}

\subsubsection{\textbf{Static taint analysis.}}
By tainting the sources in the program and analyzing the propagation of the taint tags, static taint analysis can effectively detect whether data from sources are leaked \cite{Arzt_C2014}, \cite{Gordon_C2015}. However, complex call dependency and context, and hierarchical application architecture make the static analysis more difficult and less accurate. To address these challenges, \cite{Arzt_C2014} proposes a classic information flow analysis tool based on static taint analysis, \textit{i.e.}, FlowDroid, which is now widely used for IFC through Android apps. After that, more analysis tools, such as LeakMiner \cite{Yang_C2012} and DidFail \cite{Klieber_C2014}, are proposed based on Flowdroid for more accurate analysis. Besides, new taint analysis approaches are also adopted for the data leakage detection on IoT apps.

\textit{a. Inter-component static taint analysis}. Inter-Component Communication provides a data exchange mechanism among components in Android. Analysis tools\cite{Lu_C2012}, \cite{Arzt_C2014}, \cite{Wei_C2014} mainly focus on detecting the leakage of sensitive data leaks within component. Thus Li \textit{et al.} \cite{Li_C2015} proposed IccTA (Inter-component communication Taint Analysis tool), to provide a robust and precise detection on ICC-based data leakage. IccTA enabled the connectivity of the data-flow analysis across different components by instrumenting the specified codes in apps. 

IccTA decompiled the app's Dalvik bytecodes into Jimple by Dexpler \cite{Alexandre_C2012}. Secondly, IccTA extracted the ICC links from the app's IR. And it parsed the manifest file and bytecode file to extract all defined components. All extracted information, including the ICC methods with their attribute values, were stored into a database. Then IccTA modified the Jimple representation \cite{lam_C2011} using the instrumentation approach, which directly connected the components for enabling data flow analysis between the ICC components. Three types of discontinuities were solved  including ICC methods, callback methods and lifecycle methods. Based on the revised Jimple codes, IccTA built a complete call graph of the whole Android application, which covers all ICC components. Then it performed a highly precise data flow analysis for ICC leaks detection and reported the tainted path (leakage). Based on the above design, IccTA was implemented based on Dexpler, IC3 and FlowDroid.

They conducted the experiments on a PC with Core i7 CPU PC running a Java VM with 8GB of heap size. The data sets included 22 DroidBench and 9 ICC-Bench test suites, 15000 real-world Android applications from GooglePlay, 1260 malware apps from MalGenome. Compared with four typical tools, \textit{i.e.}, FlowDroid, AppScan Source 9.0 \cite{IBM_Code2015}, DidFail, and AmAndroid, IccTA outperformed by achieving $96.6\%$ precision and $96.6\%$ recall. The peformance results showed that IccTA was almost as good as FlowDroid and better than AmAndroid.

\textit{b. Inter-language static taint analysis}. Taking efficiency and security into consideration, Android allows developers to implement a part or the complete program via native code (C/C++). It is also prone to be leveraged by attacker or malware for data stealing and evading security detection \cite{afonso_C2016} \cite{Chenxiong_C2014}. Most existing static analysis frameworks of Android such as AmAndroid \cite{Wei_C2014}, FlowDroid \cite{Arzt_C2014}, DroidSafe \cite{Gordon_C2015}, IccTA  \cite{Li_C2015} and CHEX \cite{Lu_C2012}, cannot handle native components. Wei \textit{et al.} \cite{Wei_C2018} proposed an NDK/JNI-aware inter-language framework, called JN-SAF, to track data flow across Java and native codes. JN-SAF adopted the Summary-Based Bottom-up data flow Analysis (SBDA) to efficiently perform context-sensitive inter-language data leakage detection.

Firstly, JN-SAF took apk file as input, and decompiled it into three parts, \textit{dex} files, \textit{Mainfest\&Resource} files and \textit{so} files. For Dalvik bytecode, JN-SAF leveraged DEX2IR and Resources Parser in AmAndroid to decompile them into an IR \textit{Pilar} \cite{Wei_C2014}. For binary code, JN-SAF translated binary into VEX IR\cite{shoshitaishvili_C2015} by using \textit{pyvex} from \textit{angr} \cite{Shoshitaishvili_C2016}. For resolving native method calls and activities, Wei \textit{et al.} designed a native method mapping algorithm according to the patterns on different ways of calling the native method. After that, it built environment method /function for each Java and native component, which explicitly invoked the event/lifecycle callbacks for the data flow analysis. Finally JN-SAF performed the data flow analysis on the derived environment model based on the SBDA. They leveraged Annotation-based data flow Analysis (ADA) technique to track data flow in native functions. In addition, Inter-component analyzer is built to detect ICC data flow.  Based on the above design, they implement JN-SAF based on AmAndroid and \textit{angr} tools.

They ran experiments on a PC with a 2.20GHz, 48-core Xeon CPU, and 256 GB RAM. The data set included 22 hand-crafted benchmark apps, 100,000 randomly selected popular apps from AndroZoo \cite{allix_C2016}, and 24,553 malware apps from the AMD dataset \cite{wei_C2017}. Compared with AmAndroid, Flowdroid, IccTA\cite{Li_C2015} and DroidSafe, JN-SAF outperformed all other tools in precision and recall as none has inter-language analysis capability. In performance, JN-SAF took 88.982 seconds to build native-function summary for 2,000 real-world app.

\textit{c. Static taint analysis on IoT application}. Existing IoT system lacks essential tools to support analysis on IoT apps \cite{Fernandes_C2016}. Referring to the design of FlowDroid, Celik \textit{et al.} \cite{Celik_C2018}  presented SAINT, a static taint analysis tools for IoT apps to precisely detect sensitive data flows. They carefully identified a complete set of taint sources and sinks, adequately modeled IoT app's lifecycle, and addressed platform- and language-special problems. 

During the analysis on an IoT app, an intermediate representation (IR) was first extracted from the app's source code. According to the \textit{sensor-computation-actuator} structure - a typical lifecycle of IoT apps, SAINT built the IR including Permissions block, Events/Actions block and Call graphs. Then static taint analysis was conducted on the generated IR for detecting sensitive data flows. Taint sources and sinks were defined based on analysis on platform APIs. And they applied backward taint tracking on IR for more efficient analysis. Besides, several solutions were proposed to eliminate the difficulties during taint tracking caused by the idiosyncrasies of IoT platform (SmartThings) like state variables, call by reflection, web-service IoT apps, and Groovy-specific operations. Besides, SAINT also supported tracking implicit flows. SAINT finally reported sensitive flows to users. Based on the above vital techniques, SAINT was implemented based on Groovy compiler by using \textit{Groovy AST visitors}. 
 
They applied SAINT to analyze sensitive flows in 230 IoT apps on a laptop PC with a 2.6GHz 2-core CPU and 8GB RAM. The average time per app was about 18 to 28 seconds. SAINT identified 92 out of 168 official apps and 46 out of 62 third-party apps leaking sensitive data. And there were only six extra implicit flow's warnings produced. Most flows were detected precisely with two false warnings on 19 IoTBench apps.

\subsubsection{\textbf{Static analysis based on data dependency analysis.}}
Data dependency analysis is performed on the graph generated from the application's code. SCAndroid is the first data flow analysis tool for automated security certification of Android applications\cite{fuchs_R2009}. Chex \cite{Lu_C2012} first models different entry points of Android apps  for precise flow analysis. AmAndroid \cite{Wei_C2014} improves analysis techniques by adopting more comprehensive analysis and supporting inter-component flow analysis. In recent years, more analysis approaches based on data dependency are proposed for improving analysis precision and coverage.

\textit{a. Static analysis combined with Android runtime model}. Referring to the Android application architecture in Fig.\ref{fig_app_structure}, static analysis on source package alone may cause losses on precision without consideration on the Android API and runtime semantics. Gordon \textit{et al.} \cite{Gordon_C2015} designed a usable static analysis tool, DroidSafe, which performed precisely explicit flow analysis for large, real-world Android apps. DroidSafe implemented a complete, accurate, and precise model of Android execution called the Android Device Implementation (ADI). Besides, their system adopted deep object-sensitive and flow-insensitive analysis for an accurate analysis of the Android model.

In the design of DroidSafe, ADI was firstly implemented by Java language, which which built semantic models for specific Android mechanisms including lifecycle events, callback context, external resources and inter-component communication. ADI was built on the Android Open Source Project (AOSP), version 4.0.3, \cite{Google_code}. They adopted accurate analysis stubs to modify the base code to compensate the missing runtime semantics. Then the model of inter-component communication (ICC) was built based on static communication target resolution technique. DroidSafe aggressively resolved dynamic intent values by employing Java String Analyzer (JSA)\cite{Christensen_C2003}. After that, objective sensitive points-to analysis (PTA) was performed on the Android model and Android application. DroidSafe further enhanced scalability by identifying flow-unrelated classes and eliminating object sensitivity for those classes, which enables a deeper point-to analysis. Finally, a forward data flow analysis was employed by computing over approximations of all the memory states. Based on the change of state memory, they defined a transfer function for each type of statement. Considering asynchronous callbacks without specific event ordering during run time, DroidSafe adopted a flow-insensitive analysis method for obtaining a model that correctly explored all of the event orderings. Based on the above design, they developed DroidSafe on top of the Soot Framework.

Gordon \textit{et al.} conducted the experiments on a server with an Intel Xeon CPU (E5-2690v2 3.00GHz) running Ubuntu 12.04.5 with 64GB heap memory. There were three data sets, \textit{i.e.}, a DroiDBench suite of 94 Android information-flow benchmarks, 40 Android micro applications, and 24 complete real-world Android applications. For the effectiveness, compared with FlowDroid, DroidSafe achieved the better precision and accuracy on three data sets. For the performance, the time cost by DroidSafe was higher than that of FlowDroid because DroidSafe analyzed substantially more code. Nevertheless, DroidSafe support the analysis on some large applications while FlowDroid timed out.

\textit{b. Static analysis with slice optimization}. Most static analysis tools are value-insensitive\cite{Gibler_C2012} \cite{Hoffmann_C2013} \cite{Yang_C2013} \cite{Yang_C2012} \cite{Arzt_C2014}, which may result in false alarms because of the conservative approximations for run-time values. Backes \textit{et al.} \cite{Backes_C2016} presented a novel slice-optimization tool, \textit{i.e.}, R-Droid, to extract data-dependent statements for arbitrary points within an app. It provided value analysis to simplify (semi-) automatic leak detection checks via reconstructing values/strings. 

Firstly, the application's apk file was processed for generating app's IR and constructing a comprehensive application's lifecycle model. Then R-Droid generateed an object-, context- and field-sensitive system dependency graph (SDG) by leveraging JOANA framework \cite{Hammer_J2009}. Secondly, a standard backward data-dependence slicing was performed on the SDG. It took an arbitrary list of sinks as inputs to capture data that may flow to the related statements. After that, the output was post-processed by the static slice optimization. During the optimization, they first conducted a use-def tracking analysis to eliminate spurious dependencies, then performed a precise value analysis to re-assemble strings and original values. The optimization was iteratively applied until no more slice statements were modified. Finally, these optimized slices were sent to the security check module for a broad range of security assessments, including leakage detection, user input propagation analysis, and so on. Based on the above design, R-Droid was implemented based on JOANA framework.

The experiments were conducted on a server with four Intel Xeon E5-4650L 2.60GHz CPUs and 768GB memory. The datasets included original DroidBench (v1.0) test suites and 22,700 real-world apps from Google Play. For the DroidBench set, R-Droid achieved a precision of $97\%$ and a recall of $100\%$ which was better than FlowDroid and AmAndroid's . Although DroidSafe detected all explicit flows, some sensitive flows were falsely reported. For the real-world apps set, R-Droid successfully identified 2,156 sensitive data flow in 256 apps.

\subsubsection{\textbf{Static information flow analysis based on formal modeling.}}
\textit{Sound static analysis by SMT solving}. Due to lack of soundness in most IFC approaches \cite{Enck_J2014} \cite{jiang_C2013} \cite{Arzt_C2014} \cite{Gordon_C2015} \cite{Li_C2015}, Calzavara \textit{et al.} \cite{Calzavara_C2016} presented a new data flow analysis based on formal modeling, \textit{i.e., Horn clause resolution} \cite{bjorner_J2012}. The formal analysis-based approach soundly abstracted the semantics of Android apps. And security properties were formulated as a set of proof obligations. Then it automatically verified those properties through off-the-shelf SMT solvers. During the analysis, the soundness was guaranteed by their formal proof.

They first designed a rigorous formal model of the Android semantics by extending the $\mu$-Dalvik calculus \cite{jeon_R2012}, \textit{i.e.}, $\mu$-Dalvik$_A$. In their model, they defined $\mu$-Dalvik$_A$'s syntax, the operational semantics and activity semantics, the activity lifecycle, and inter-component communication. Then they translated an input program into a abstract model based on $\mu$-Dalvik$_A$'s semantics, which was represented by a set of Horn clauses \cite{bjorner_J2012}. After that, they proposed a Preservation theorem which stated that any reachable configuration was over approximated by some set of facts in Horn clauses. Finally, they performed a sound static data flow analysis to detect the undesired information flows according to the theorem. During the analysis, queries on the sensitive flows were automatically generated based on a public-available database of sources and sinks \cite{rasthofer_C2014}. Based on the above theory, Calzavara \textit{et al.} developed HornDroid, the first static data flow analysis tool for Android with a formal proof of soundness, based on the Z3 SMT solver \cite{Moura_J2008}.

They firstly conducted an evaluation on Droidbench. Compared with IccTA, AmAndroid, DroidSafe, HornDroid outperformed in terms of both soundness and performance with a high precision. HornDroid was the only tool that identified all 96 applications containing explicit leaky flows. For the performance, HornDroid was much faster than IccTA, AmAndroid, and DroidSafe. For Candy Crash application, it appeared to be secure through the analysis of HornDroid. AmAndroid got an information flow which is proved to be false alarm. Neither IccTA nor DroidSafe was not able to analyze the app within 2 hours.

\subsubsection{\textbf{Semantic-based static information flow analysis.}}
\begin{figure}[t]
\setlength{\abovecaptionskip}{0.2cm}
\centerline{\includegraphics[scale=.4]{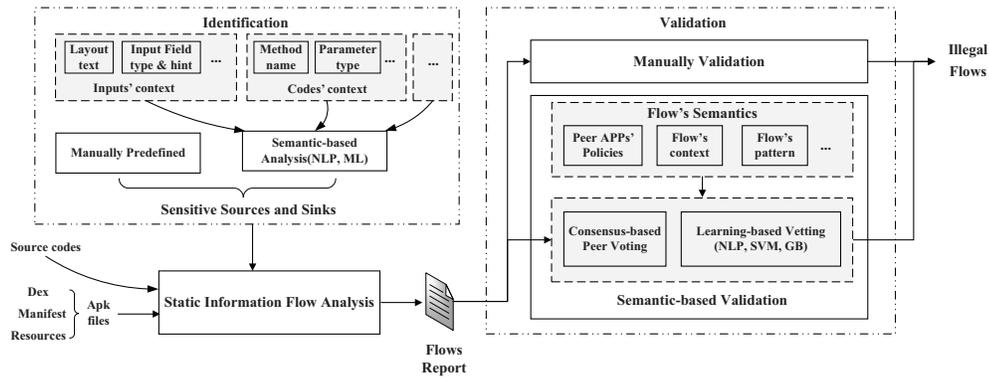}}
\caption{Semantic-based Validation Approaches}
\label{fig_sem_validation}
\vspace{-0.4cm}
\end{figure}

Above approaches provides various precise tracking techniques, while the legality of the detected flows is not discussed. With the development of ML (machine learning), semantic-based IFC techniques are widely studied in recent years. \cite{rasthofer_C2014} firstly designed a novel and fully automated machine-learning tool, SUSI, for identifying Android app's sources and sinks according to their syntactic and semantic features. SUSI can identify hundreds of sensitive flows missed by previous information-flow tracking tools. Except for sources and sinks, more semantic features were employed in identifying illegal information flows in recent years. The general procedure of these works are shown in Fig.\ref{fig_sem_validation}. 

Based on the results of the static information flow analysis procedure, these works introduce the precise identification on sensitive flows\cite{Nan_C2015} \cite{Huang_C2015} \cite{Pan_C2018} \cite{Nan_C2018}, and implement automatic validation on detected flows \cite{Vitalii_C2015} \cite{Lu_C2015}. Five representative works are summarized in Table \ref{table_semantic_analysis} in our survey. But these mainly focus on Android app. Semantic-based information flow analysis on IoT app is still missing. We compare them from six aspects, including research goal, analyzed semantics, analysis objects, approach type, and main techniques. Then we illustrate the evaluations in Table \ref{table_eva_semantic_analysis}. The basic settings include the environment, capability, and data set. The evaluation results include the total number of tested apps or flows, the precision, the recall, the accuracy and the performance.

\textit{a. Precise sensitive user input detection.} During flow verification, sensitive and non-sensitive user inputs are usually mixed and submitted to the application. Neglection may cause the miss of data leakage while conservative protection could result in false alarms. To overcome these flaws, Huang \textit{et al.} \cite{Huang_C2015} proposed a mobile app analysis tool called SUPOR (Sensitive User inPut DetectOR) to precisely identify sensitive user inputs. They leveraged UI rendering, geometrical layout analysis and NLP techniques. They also designed a context-sensitive approach to bind sensitive UI input fields to the corresponding variables in the code. 

SUPOR consisted of three major components, \textit{i.e.,} layout analysis, UI sensitiveness analysis and variable binding. Firstly, the layout analysis component accepted an APK file as the input. It parsed the layout files for identifying all input fields (e.g., \textit{EditText}). Then the coordination information for the input fields was obtained by rendering the layout files. SUPOR also collected text labels and their attributes of input fields, such as \textit{ID} and \textit{inputType}. Then the UI sensitiveness analysis component judged the sensitivities of input fields. Input field's attributes, hint and neighbor text label were checked according to their sensitive keyword dataset which was extracted from the UIs of a tremendous number of apps by leveraging NLP. After that, the variable binding component built the mapping relation between sensitive input fields and the variables in the code by using a context-sensitive analysis.. Finally, these identified variables were labelled as sensitive sources for static or dynamic analysis. According to the above design, they implemented SUPOR based on Apktool, dexlib and WALA framework, UI rendering engine, and Dalysis. 

The experiments were conducted on a cluster of eight servers,  with server equipped with an Intel Xeon E5-1650 CPU and 64/128GB of RAM. The data set included 16000 apps from the Google Play store. For the performance,  11.1 apps were analyzed per minute on the eight-server cluster. SUPOR had only one false negative in 20 apps without sensitive input fields. SUPOR identified 149 sensitive input fields with 4 FPs (False Positive) and 3 FNs (False Negative) in 20 apps with sensitive input fields. SUPOR achieved an accuracy of $91.3\%$ in 4922 real-world apps.

\begin{table}[t]
\setlength{\abovecaptionskip}{0.2cm}
\footnotesize
\newcommand{\tabincell}[2]{\begin{tabular}{@{}#1@{}}#2\end{tabular}}
\centering
\caption{Semantic-based Information Flow Validation Approaches}
 \resizebox{\textwidth}{19mm}{
\begin{tabular}{ccm{3cm}m{2cm}ccm{3cm}}
\toprule
\textbf{Work} & \textbf{Platform} &  \makebox[3cm]{\textbf{Goal}}&  \makebox[2cm]{\textbf{Semantics}} &  \textbf{Analysis Objects} & \textbf{Approach} & \makebox[3cm]{\textbf{Techniques}}  \\
\midrule
\textbf{SUPOR} \cite{Huang_C2015}  & Android & Identification on sensitive inputs & User's inputs & Layouts & Keyword-based & UI rendering, NLP \\
\specialrule{0em}{2pt}{1pt}
\textbf{FlowCog} \cite{Pan_C2018}  & Android & Validation on Flows & Flow's contexts & Flow-related Views & Learning-based & NLP, SVM, Gradient Boosting (GB)\\
\specialrule{0em}{2pt}{1pt}
\textbf{ClueFinder} \cite{Nan_C2018}  & Android & Identification on sensitive sources & Code statements & Codes & Learning-based & NLP, SVM \\
\specialrule{0em}{2pt}{1pt}
\textbf{AAPL} \cite{Lu_C2015}  & Android  & Validation on Flows & Similar APPs' decisions & Similar Apps & Consensus-based  & Peer Voting \\
\specialrule{0em}{2pt}{1pt}
\textbf{MUDFLOW} \cite{Vitalii_C2015}  & Android & Validation on Flows & Flow's pattern & Detected Flows & Learning-based & SVM \\
\bottomrule
\end{tabular}}
\label{table_semantic_analysis}
\vspace{-0.3cm}
\end{table}

\textit{b. Context-aware semantics analysis of information flows.} Most existing approaches \cite{Arzt_C2014}, \cite{Gordon_C2015}, \cite{Enck_J2014}, \cite{Klieber_C2014}, \cite{Li_C2015} are mainly staying on code analysis level with less considering the semantics of the code contexts. Therefore, they may cause a high false-negative rate on data leakage detection. For improving the accuracy,  Pan \textit{et al.} \cite{Pan_C2018} proposed an automated, semantic-sensitive flow analysis system, called FlowCog. The system extracted semantics from each information flow, called context, in an Android app. Based on the flow-specific context, FlowCogs could automatically check whether the flow is legitimate precisely. 

FlowCog firstly found all the information flows of the app by Flowdroid. Then the system focused on two special statements, i.e., \textit{activation event} and \textit{guarding condition}. These two special statements influenced the user's decision and perception on the data flow. Special Statement Discovery Engine extracted them by static analysis approach. Next, it used data flow analysis technique to find the view dependency between views and data flow. After that, FlowCog extracted contexts for each flow based on the depended views, app's descriptions and view's layout. For the semantics on app's descriptions, FlowCog used existing approaches \cite{Gorla_C2014}, \cite{Qu_C2014} to extract them. For the semantics on depended views, the system extracted the semantics-related texts from depended views or other views in the same layout. Finally, FlowCog leverageed NLP approach to associate a given flow with corresponding extracted semantics. Based on the output of the classifier, the system could determine whether the flow was negative. Pan \textit{et al.} implemented FlowCog based on FlowDroid, Soot, and Stanford parser.

Their experimental platform contained a 2.8 GHZ Intel Xeon CPU and 32G memory, running Ubuntu 14.04 LTS. The overall dataset contained 6,000 benign and malicious apps, in which 4,500 benign apps were randomly crawled from Google Play, and 1,500 malicious ones were randomly selected from Drebin dataset\cite{arp_C2014}. The precision, recall and accuracy of FlowCog are 90.1$\%$, 93.1$\%$ and 90.2$\%$ respectively. The results showed that flow contexts could facilitate the detection, \textit{i.e.}, $10\%$ improvement in terms of accuracy, compared to app-level semantics (app's description) alone.

\textit{c. Semantics-Driven and learning-based leakage detection.} Besides the flow's context, Nan \textit{et al.} \cite{Nan_C2018} observed that app's code contains a large amount of semantic information to explain meaningful program elements. Leveraging this observation, they developed a semantics-driven static analysis tool, namely ClueFinder, for automatically and precisely locating sensitive sources in Android apps. They utilized NLP to automatically identify program tokens, such as variables and methods. Then, they performed a learning-based approach to analyze the program structure and accurately identify sensitive content.

Firstly, all program elements related to privacy were selected according to the program's semantics. They defined a set of data considered to be sensitive and key words associated with them, based on Google Privacy Policies and prior privacy-related research \cite{Nan_C2015}. Additionally, they adopted NLP techniques to find out the raw target elements, \textit{i.e., code fragment}. Then semantics checker refined these raw elements to filter out false-positive data by Parts-Of-Speech (POS) tagging and dependency analysis approach. After that, the rest of the elements were processed by structure analyzer for more precise identification. They utilized a SVM classifier to determine whether a related statement involved with private data. The classifier was based on features of \textit{method name, parameter type, return type, base value type} and \textit{constant-variable pattern}. Finally, Cluefinder used static taint analysis techniques to track the data objects in sensitive statements. Based on the proposed approach, they implemented ClueFinder in Java and Python based on FlowDroid, Stanford Parser \cite{Klein_C2003} and the SVM \cite{Scikit_code} classifier.

They conducted the experiments on a 32-core CPU server and 64GB memory. ClueFinder achieved a precision of 92.7$\%$, a recall of 97.2$\%$ and an F1-Score of 94.8$\%$. it took less than 1 minute to analyze each app.  In addition, they compared ClueFinder with SUSI \cite{rasthofer_C2014}, UIPicker\cite{Nan_C2015}, SUPOR \cite{Huang_C2015}, and BidText\cite{Huang_C2016}. Over 77.6$\%$ of statements were found to be false positives by using SUSI. ClueFinder identified all the UI sources in most cases, and also found two times more non-UI sources missed by UIPicker and SUPOR. Compared with BidText, ClueFinder reduced the false positive rate by 6\%  and executed 1.86 times faster.

\begin{table}[t]
\setlength{\abovecaptionskip}{0.2cm}
\footnotesize
\centering
\caption{Evaluations on Semantic-based Information Flow Validation Approaches}
\resizebox{\textwidth}{21mm}{
\begin{tabular}{ccm{2cm}<{\centering}m{2cm}<{\centering}m{2cm}<{\centering}m{1cm}<{\centering}cccm{1cm}<{\centering}}
\toprule
\textbf{Work} & \textbf{Environment} &  \textbf{Capability} &  \textbf{Data Set} & \textbf{$\#$ App/Flow} & \textbf{Precision}  & \textbf{Recall} & \textbf{Accuracy} & \textbf{Performance (s/app)} \\
\midrule
\textbf{SUPOR} \cite{Huang_C2015}  & Server, JVM & Xeon E5-1650, 16GB heap size & GooglePlay & 20 & 97.3$\%$ & 97.3$\%$ & - & 5.2\\
\specialrule{0em}{2pt}{1pt}
\textbf{FlowCog} \cite{Pan_C2018}  & Server & 16 core, 32GB memory & GooglePlay and Drebin & 6,000(P$^1$)+1,500(N$^2$) & 90.1$\%$ & 93.1$\%$ & - & - \\
\specialrule{0em}{2pt}{1pt}
\textbf{ClueFinder} \cite{Nan_C2018}  & Server & 32 core, 64GB memory & Manually Labeled Flows & 2,316(P)+2,316(N) & 92.7$\%$ & 97.2$\%$ &- & 55  \\
\specialrule{0em}{2pt}{1pt}
\textbf{AAPL} \cite{Lu_C2015}  & Server  & Xeon E5-1650, 64GB memory & Labeled Flows & 266 & - & - & 88.7$\%$ & -\\
\specialrule{0em}{2pt}{1pt}
\textbf{MUDFLOW} \cite{Vitalii_C2015}  & Server & 64 core, 730GB memory & GooglePlay, Genome and VirusShare & 2,866(P)+15,338(N) & - & - & 83.8$\%$ & [1, 100000]  \\
\bottomrule\\
\multicolumn{9}{c}{$1$: Accuracy = TP+TN/(TP+FP+TN+FN)}\\
\multicolumn{9}{c}{$2$: P = Positive/Benign, $3$: N = Negative/Malicious}
\end{tabular}}
\label{table_eva_semantic_analysis}
\vspace{-0.4cm}
\end{table}

\textit{d. Data leakage detection based on peer voting.} Current data-flow analysis techniques \cite{Gibler_C2012} \cite{Lu_C2012} \cite{Enck_J2014} \cite{Rastogi_C2013} have difficulties on identifying complex flows including conditional and joint flows in Android apps. Besides, single detection result contain too many false positives. Lu \textit{et al.} \cite{Lu_C2015} proposed a scheme named as AAPL (Analysis of App Privacy Leak) which had the capability to check the complex flows and judging the legitimacy of privacy disclosures. 

Based on the idea of judging legitimacy by peer vetting mechanism, the primary app and its functionally similar peer apps were firstly found by the recommendations from the market, and purified by NLP technique \cite{resnik_J1999}. Then, an enhanced information flow analysis was conducted on the primary app and its peer apps. They designed three new techniques, opportunistic constant evaluation, object origin inference and joint flow tracking, to accurately obtain all privacy disclosure flows in the primary app and its peer apps. After that, they proposed a new peer app voting mechanism to differentiate the legal disclosure flows from the suspicious ones automatically. Based on the voting results, AAPL computed the privacy legitimacy and reported the result to the user. AAPL's privacy disclosure module was built on top of \textit{Dalysis} \cite{Lu_C2012} and IBM \textit{WALA}. 

The experiments were conducted on a cluster of three servers, with server having a Xeon E5-1650 CPU and 64 GB memory. They analyzed 40,506 apps (primary apps and peer apps) in total and found 44.7$\%$ have privacy disclosures. Compared to the traditional analysis, the detection rate increased by 11.5$\%$. Furthermore, the average analysis time was about 4.5 apps per minute. For the peer voting mechanism, AAPL achieved an average accuracy of 88.7$\%$, 10.7$\%$ FP rate and 12.5$\%$ FN rate. 

\textit{e. Flow's pattern-based abnormal usage detection.} Besides above semantic features, Vitalii \textit{et al.} \cite{Vitalii_C2015} built a tool called MUDFLOW (Mining Unusual Data Flow) by leveraging benign flow's pattern. MUDFLOW adopted multiple one-class classifiers, training on the sensitive data flow usage of benign apps, to automatically identify malicious apps. 

MUDFLOW firstly used FlowDroid to extract all data flows from sensitive data sources to sinks within one application. MUDFLOW aggregated these flows as Category-level flows according to SUSI's category of the API \cite{rasthofer_C2014}. After that, MUDFLOW took a set of begin apps as a ground truth set, then identified which apps had unusual flows within each category. For each category, MUDFLOW computed the distance between the new app and its k-nearest neighbours in begin app set, which served as an outlier score in this category. Then MUDFLOW applied the above outlier detection on every SUSI source category and obtained an aggregated score vector, \textit{i.e.}, a vector of distances with different weights to different categories for emphasizing different importance. Finally, MUDFLOW leveraged \textit{v}-SVM one-class classification \cite{Chen_J2005} to determine whether an app is malicious. The training set was constructed based on benign applications, and their score vectors were used as features. Based on the above design, MUDFLOW was built on FlowDroid and the implementation of multiple classifiers.

The experiment were conducted on a server with 730 GB of RAM and a 64-core Intel Xeon CPU. There were a total of 2950 apps from Google Play Store as their initial benign data set. Malicious apps set included 1260 malware apps from the Genome project and 24,317 malicious ones from VirusShare database. They executed 10-fold cross validation on two data sets. The results showed that MUDFLOW identified $90.1\%$ of the malware with a false positive rate of $18.7\%$. For the collected 96 repackaged apps, MUDFLOW identified $97.6\%$ of them. 

\subsection{Run-time information flow control}
Deploying-time IFC mechanisms provide an effective way but with precision loss due to the conservative analysis of complex objects or methods \cite{Li_C2015} \cite{Calzavara_C2016}, approximate runtime model \cite{Gordon_C2015}, \textit{etc}. To overcome these flaws, run-time IFC mechanisms based on dynamic analysis are also well studied. During the run-time analysis, application's executions, including instructions, memory, register, \textit{etc.}, are monitored to capture or block the data leakage flows precisely and timely. But it requires modifications on applications, runtime framework, system and even hardware. Due to the extra overhead on execution, run-time IFC can be executed under emulated environments sometimes. 

Works on run-time information flow control approaches are categorized into 3 groups according to different techniques, \textit{i.e.}, dynamic taint analysis, secure multi execution (SME), and type system. These approaches are summarized in Table \ref{table_runtime_control}. We compare recent 11 representative works from eight metrics including \textit{platform}, \textit{techniques}, \textit{modification level}, \textit{security levels}, \textit{ICC}, \textit{IAC}, \textit{implicit flow}, and \textit{soundness}. For security levels, it is used to represent data with different sensitivities which are introduced by the multilevel security model. For IAC, it means whether this work supports the analysis on inter-application communication.  Besides, we also make a comparison on the evaluation settings and results in Table \ref{table_eva_runtime_control}. The evaluation settings include test environment and data set. For evaluation results, we focus on the effectiveness and performance of different run-time approaches. For effectiveness, it mainly includes the capability of detection on illegal apps/flows, compatibility with the low-level system, or resistance to specific attacks. For performance, we focus on the extra overhead on the application's execution due to the modifications on applications. Main contributions of each work are detailed as follows. 

\begin{table}[t]
\setlength{\abovecaptionskip}{0.2cm}
\footnotesize
\newcommand{\tabincell}[2]{\begin{tabular}{@{}#1@{}}#2\end{tabular}}
\centering
\caption{Run-time information flow control Approaches}
\resizebox{\textwidth}{32mm}{
\begin{tabular}{ccp{3cm}p{3cm}ccccc}
\toprule
\textbf{Work} & \textbf{Platform}  &  \makebox[3cm]{\textbf{Techniques}}&  \makebox[3cm]{\textbf{Modification Level}} &  \textbf{Security Levels$^1$} & \textbf{ICC} & \textbf{IAC$^2$}  & \textbf{Implicit Flow} & \textbf{Soundness}\\
\midrule
\textbf{TaintART} \cite{Sun_C2016}  & Android & Dynamic Taint Analysis & Application, Middleware & Multi Levels & $\checkmark$ & $\times$  & $\times$ &  $\times$ \\
\specialrule{0em}{2pt}{1pt}
\textbf{TaintMan} \cite{You_J2017}  & Android & Hybird Analysis & Application, Middleware &  Multi Levels & $\checkmark$ & $\times$ & $\bullet$ & $\bullet$ \\
\specialrule{0em}{2pt}{1pt}
\textbf{PIFT} \cite{Yoon_C2016}  & Android & Dynamic Taint Analysis & Application, Middleware, System and Hardware & Two Levels & $\checkmark$ & $\times$ & $\times$ & $\times$ \\
\specialrule{0em}{2pt}{1pt}
\textbf{CDroid} \cite{Wu_J2018}  & Android & Dynamic Taint Analysis & Application, Middleware & Multi Levels & $\checkmark$ & $\checkmark$ & $\times$ & $\times$ \\
\specialrule{0em}{2pt}{1pt}
\textbf{NDroid} \cite{Xue_J2019} & Android & Hybird Analysis & Application, Middleware, System & Multi Levels & $\checkmark$ & $\times$ & $\times$ \\
\specialrule{0em}{2pt}{1pt}
\textbf{ContexIoT} \cite{Jia_C2017}  & IoT & Hybird Analysis & Application & Multi levels & - & $\checkmark$ & $\checkmark$ & $\times$  \\
\specialrule{0em}{2pt}{1pt}
\textbf{FlowFence} \cite{Fernandes_C2016}  & IoT & Quarantined Execution, Dynamic Taint Analysis & Application, Middleware & Multi Levels & - & $\checkmark$ & $\checkmark$ & -\\
\specialrule{0em}{2pt}{1pt}
\midrule
\specialrule{0em}{2pt}{1pt}
\textbf{Weir} \cite{Nadkarni_C2016}  & Android & Secure Multi Execution & System & Multi Levels & $\checkmark$ & $\checkmark$ & $\checkmark$ & $\checkmark$ \\
\specialrule{0em}{2pt}{1pt}
\textbf{Ariel} \cite{Chakraborty_C2019}  & Android & Secure Multi Execution & System & Two Levels & $\checkmark$ & $\checkmark$ & $\checkmark$ & $\checkmark$ \\
\specialrule{0em}{2pt}{1pt}
\midrule
\specialrule{0em}{2pt}{1pt}
\textbf{FlowIT} \cite{Bastys_C2018}  & IoT & Type System & Application & Two Levels & - & $\checkmark$ & $\checkmark$ & $\checkmark$ \\
\specialrule{0em}{2pt}{1pt}
\textbf{CaITApp} \cite{Balliu_C2019}  & IoT & Type System & Application & Multi Levels & - & $\checkmark$ & $\checkmark$ & $\checkmark$\\
\bottomrule\\
\multicolumn{9}{c}{$1$: Security Levels: Supported security levels in Multilevel security}\\
\multicolumn{9}{c}{$2$: IAC = Inter-Application Communication}
\end{tabular}}
\label{table_runtime_control}
\vspace{-0.4cm}
\end{table}

\subsubsection{\textbf{Dynamic taint analysis}} Dynamic taint analysis (DTA) is one of the most representative techniques used in mobile and IoT app's vulnerabilities and malware detection. \cite{Enck_C2010} proposed the first dynamic taint analysis system, \textit{i.e.,} TaintDroid,  to achieve practical system-wide information flow analysis in Android. Many systems \cite{Dam_C2012} \cite{Balebako_C2013} \cite{Rastogi_C2013} \cite{Qian_C2014} conducted further analysis based on TaintDroid. \cite{Kwong_C2012} realizes tracking information leakage through both Java and native components in an Android app. In recent years, more dynamic taint analysis are proposed for improvements. We detail six representative works as follows \cite{Sun_C2016} \cite{You_J2017} \cite{Yoon_C2016} \cite{Wu_J2018} \cite{Xue_J2019}  \cite{Jia_C2017}.

\textit{a. Dynamic taint analysis in ART.} Although TaintDroid \cite{Enck_C2010} \cite{Enck_J2014} is infeasible in newer versions of Android with Android RunTime environment (ART). Sun \textit{et al.}\cite{Sun_C2016} designed a multi-level information flow tracking system for ART, called TaintART. TaintART overcame the compatibility and performance issues of Taintdroid. They conducted multi-level taint analysis on optimized apps rather than the original one with bytecode. They achieved fast taint storage and propagation for performance improvement by using processor registers. 

TaintART first utilized processor registers to store taint tags on different variables for efficiency. Main memory was used for temporary storage of extra variables.  For scalability, TaintART supported multi-level taint tracking by extending the bits of taint tags in the register. Then they designed new taint propagation logics including basic propagation, propagation via methods calls, and propagation between apps through binder IPC (Inter Procedural Communication). After that, TaintART inserted taint propagation code blocks into app's native code during the app's compiling in ART. The logic was implemented on the internal instruction (HInstruction). HInstruction was generated by parsing and optimizing the application's dex bytecode. Finally, TaintART tracked taint tags of sensitive data efficiently at run time. Once the tainted data was transferred to illegal channels, TaintART immediately reported the incident. Based on the above design, they implemented TatinART by customizing the ART compiler, ART runtime sources, and Android framework sources.

They conducted experiments on a Nexus 5 device. Operating System was Android 6.0. Their test data sets included self-developed applications, 80 built-in apps in AOSP (Android Open Source Project), and benchmarks in CaffeineMark 3.0 \cite{CaffeineMark_Code}.The result showed that TaintART could successfully capture the leakage in popular apps without compatibility issues. In addition, TaintART incurred less than 15$\%$ Java runtime overhead compared to the original environment. TaintART also incurred little memory overhead and less than 5$\%$ IPC overhead. 

\begin{table}[t]
\setlength{\abovecaptionskip}{0.2cm}
\footnotesize
\newcommand{\tabincell}[2]{\begin{tabular}{@{}#1@{}}#2\end{tabular}}
\centering
\caption{Evaluations on Run-time information flow control Approaches}
\resizebox{\textwidth}{40mm}{
\begin{tabular}{cm{3.5cm}m{4cm}m{4.5cm}m{4cm}}
\toprule
\textbf{Work} & \makebox[3.5cm]{\textbf{Environment}} & \makebox[4cm]{\textbf{Data Set}} & \makebox[4.5cm]{\textbf{Effectiveness}} & \makebox[4cm]{\textbf{Performance}} \\
\midrule
\textbf{TaintART} \cite{Sun_C2016}  & Nexus 5, Quadcore 2.3 GHz CPU, 2GB memory, Android 6.0.1 & Popular apps (Taobao, Facebook, etc.); 80 built-in apps in AOSP; and benchmarks in CaffeineMark 3.0 & Capture the leakage in popular apps. Analyze apps without compatibility issues. & Less than 15$\%$ run-time overhead, negligible memory overhead and less than 5$\%$ IPC overhead. \\
\specialrule{0em}{2pt}{1pt}
\textbf{TaintMan} \cite{You_J2017}  & HTC One with Android 4.0, MOTO G with Android 5.0 & 150 malware samples in Genome;  100  popular apps; 9 proof-of-concepts and 2 real-world malware  & Detect all malware samples without false alarm. Find 4 more apps containing leakage compared with TaintDroid. Detect the implicit-flow leakage. & 42.3$\%$performance overhead and 28.9$\%$ with optimization. \\
\specialrule{0em}{2pt}{1pt}
\textbf{PIFT} \cite{Yoon_C2016} & Gem5 simulator with Android 4.2 & 57 DriodBench apps; 7 real-world malwares & Achieve 98$\%$ accuracy without false positive. & Limited impact on performance and added CPU complexity \\
\specialrule{0em}{2pt}{1pt}
\textbf{CDroid} \cite{Wu_J2018}  & Laptop, i5 2.4 GHz, 4 GB RAM, Android 4.1.1 & WPS application; benchmarks in Benchmark and CaffeineMark & Block sending out sensitive messages in WPS. & 5$\%$ extra memory usage and 17$\%$ runtime overhead. \\
\specialrule{0em}{2pt}{1pt}
\textbf{NDroid} \cite{Xue_J2019} & Android 4.3 and Android 5.0 & 2 real-world apps (QQPhoneBook and ePhone); two real-world malwares (SpyBubble and PlusLock), and CaffeineMark benchmark. & Track the information leakage through  multiple contexts. & 10.7 and 10.1 times slowdown to the DVM runtime and ART runtime respectively. \\
\specialrule{0em}{2pt}{1pt}
\textbf{ContexIoT} \cite{Jia_C2017} & SmartThings platform & 25 self-developed SmartApps  and  283 commodity SmartApps & Identify all attacks to 25 apps without ambiguity. & 3.5 times permission request frequency and 67.1 ms additional delay. \\
\specialrule{0em}{2pt}{1pt}
\textbf{FlowFence} \cite{Fernandes_C2016}  & LG Nexus 4, IoT hub with Android 5.0 & Microbenchmarks, 3 typical IoT apps (SmartLights, FaceDoor, HeartRate Monitor) & Eliminate possible insecure flows & 99 lines code and 4.9$\%$ latency increment.\\
\specialrule{0em}{2pt}{1pt}
\textbf{Weir} \cite{Nadkarni_C2016} & Nexus 5, Android 5.0.1 & 30 top apps from GooglePlay & Polyinstantiation without compatibility issues. & Less than 4ms overhead for starting components, 5.98 ms for copy operation, and 2.02 ms for establish a network connection. \\
\specialrule{0em}{2pt}{1pt}
\textbf{Ariel} \cite{Chakraborty_C2019}  &  Android 6.0 & 2 case studies (permission re-delegation attacks and untrusted third party code) & Mitigate these two well-known attacks & - \\
\specialrule{0em}{2pt}{1pt}
\textbf{FlowIT} \cite{Bastys_C2018}  & - & 60 IFTTT applets (30 secure +30 insecure) & No false negatives, and 1 false positive. & -  \\

\specialrule{0em}{2pt}{1pt}
\textbf{CaITApp} \cite{Balliu_C2019}   & - &  Four examples inspired by real-world apps & All could be modeled in their calculus & -  \\
\bottomrule
\end{tabular}}
\label{table_eva_runtime_control}
\vspace{-0.4cm}
\end{table}

\textit{b. Hybird taint analysis in ART.} Sophisticated malware may evade the detection in emulator-based environment \cite{Kwong_C2012} by employing anti-analysis techniques. And most DTA approaches failed to detect implicit flows caused by control dependence. To address these challenges, You \textit{et al.} \cite{You_J2017} designed and implemented a hybird dynamic taint analysis tool, \textit{i.e.}, TaintMan, to support detection on explicit and implicit flows in ART. By combining with static analysis technique, they tracked strict control dependencies in Android apps. They also proposed a novel technique called \textit{reference hijacking} to provide a more precise analysis without rooting the device. Furthermore, they enforced on-demand instrumentation and tracking to optimize the performance on real devices. 

TaintMan first performed static instrumentation on both target app and its related system libraries. The taint enforcement codes, including taint storage and propagation, were added into the original bytecode files, \textit{apk file}. Besides, they designed lazy tainting rules based on static analysis for  a specific implicit flow tracking, \textit{i.e.}, strict control dependence (SCD). Then, they modified the manifest file to alter the entry point of target  for the preparation on reference hijacking. These modified files are packed with the original resource files together. Finally, the instrumented application ran on mobile devices with a customized execution environment constructed by reference hijacking. In this environment, both the target app and system libraries could be dynamically tracked without rooting. During the application's execution, the system reported an alert when tainted data went out through an illegal sink method. Besides, TaintMan adopted on-demand instrumentation nd tracking to reduce the overhead cost.

They deployed TaintMan on two types of smartphones, HTC One S equipped with Android-4.0 and Motorola MOTO G equipped with Android-5.0. The data sets included 150 malware samples form the Android Malware Genome Project, 100 popular apps from multiple markets, nine proofs-of-concept from their prior work, and two real-world malware samples. TaintMan could detect the data leakage in all 150 malware samples, and none of them is a false alarm. For the test on 100 real-world apps, TaintMan performed better than TaintDroid on the leakage including implicit flows. The evaluation also showed that TaintMan incurred 42.3\% performance overhead without optimization and 28.9\% with optimization. But the size of app increased about 23\%, and the instrumented system libraries were nearly three times as large as the original ones.

\textit{c. Predictive information-flow tracking.} Dynamic analysis approaches usually come with significant run-time overhead due to the additional tracking work. Yoon \textit{et al.} \cite{Yoon_C2016} proposed a predictive information flow tracking approach (PIFT) to reduce the cost. PIFT achieved better performance by simplifying the complex and high-cost mechanics of tracking. The PIFT approach was built on the observation and statistical analysis on the Android system's memory load and store instructions. The observation showed that most computations consisted of short-lived, possibly interleaved \textit{load-process-store} sequences.

PIFT was a four-level taint tracking system including PIFT Manager, PIFT Native, PIFT Software and Hardware from top to bottom. It corresponds to four levels in Android's structure in figure \ref{fig_app_structure}. When the object instance fetched the sensitive data through specific sources,  it was registered and passed down by PIFT Manager for the following tracking (tainted). When the object instance was sent out by specific sinks, it was passed to the lower layer for checking. Then the object instance was translated by PIFT Native to its corresponding memory addresses. After that, the addresses were passed down to PIFT Software and Hardware Module for executing the taint propagation. Instead of precisely tracking each specific operation at register-level, PIFT's taint propagation algorithm tainted multiple memory ranges located in \textit{Tainting Window}, \textit{i.e., the target addresses} of the next few store instructions. When detecting taint locating in the checked address range, it informed the upper layer about the potential leakage. For efficiency, PIFT Hardware Module adopted cache-based taint storage for quick lookup operations.

They ran experiments on gem5 simulator \cite{Binkert_C2011} with Android 4.2 on ARM processor (RISC-style load/store architecture). The data set included 57 DriodBench apps (41 leaky and 16 benign). The results showed that PIFT could achieve $98\%$ accuracy and 0 false positives when they selected appropriate window size. They also tested seven real-world malware apps. Leakage flows were all detected by PIFT with limited impact on performance and CPU complexity.

\textit{d. Centralized information flow control in Android.} Android is full of untrusted third-party apps. It is risky to let the app control users' private data. Wu \textit{et al.} \cite{Wu_J2018} proposed a novel centralized information flow control (CIFC) model, which could control private data by the user rather than apps. They constructed the CIFC formal model using a Value-passing Security Process Algebra (VSPA), which was proved to be able to guarantee the noninterference security property. By combining with DTA, CDroid realized tracking information flow at run time under users' monitoring. 

The CIFC model was designed first based on VSPA, a formal modeling language. The model consisted of subject, object, reference monitor, policy configuration, audit and declassification. In the CIFC model, all information flows were monitored and checked by reference monitor. The noninterference security was proved by Checker of Persistent Security (CoPS) tool \cite{Carla_C2004}. After that, they implemented a prototype system called CDroid by modifying the application layer and the middleware layer. In the application layer, they provided three new applications including of \textit{CPolicy}, \textit{CNotify} and \textit{FAdmin}. These applications were used to manage flow policies and notify the violations. In the middleware layer, They used TaintDroid \cite{Enck_J2014} to realize dynamic taint tracking. They also deployed optimization methods on hot traces for providing light-weight taint propagation analysis.

Experiments and evaluations were performed on Android emulator. The version of Android was 4.1.1$\_$r6. CDroid successfully distinguished the legal and illegal behaviours in three real apps. The results also showed that CDroid caused 5$\%$ additional memory usage and 17$\%$ runtime overhead compared with the original Android. However the centralized management of flow policies was a complex and tedious work, which may put more burden on users. 

\textit{e. Tracking information flows across multiple Android contexts.} Many Android apps use native code and interface with Java code. However, this interaction causes many troubles to existing dynamic analysis systems. Some DTA tools, including TaintDroid \cite{Enck_J2014} and TaintART \cite{Sun_C2016}, could cause under-tainting\cite{Kang_C2011}. The cost is also important. DroidScope \cite{Kwong_C2012} can track the entire system’s data flow, but the cost is very high. To address the above problems, Xue \textit{et al.} \cite{Xue_J2019} proposed NDroid based on DTA techniques. NDroid can trace data leakage from multiple contexts (such as Java context and native context. By modifying the Android kernel and application framework, they instrumented instructions at the compilation phase and tracked sensitive information flows at the execution phase with the help of QEMU \cite{Qemu_Code}. 

NDroid introduced six major modules into QEMU for DVM and ART to track the flows, including instrumentation manager, taint engine, JNI tracer, instruction tracer, native tracker and DVM/ART tracker. Firstly, the instrumentation manager was introduced to do instrumentation on different modules. These modules included JNI bridge, system libraries and native libraries. The instrumentation included two different levels, basic block level and instruction level. Then dynamic information flow tracking was performed by cooperations among JNI tracer, instruction tracer, native tracker and DVM/ART tracker. The JNI tracer performed the taint propagation on JNI APIs and related functions. These functions mainly included \textit{JNI entry}, \textit{method calls}, \textit{object/string/array operation}, \textit{field access} and \textit{exception}. The instruction tracer traced information flows in native codes by analyzing the instrumented instructions. The native tracer conducted the taint propagation on the compiled code. During the tracing procedure, the taint engine maintained a taint map and shadow registers to store taint status of memory and register, respectively.

The experiments were conducted on Android 4.3 with DVM runtime environment and Android 5.0 with ART environment. The data sets included two real-world apps, two malwares, and CaffeineMark \cite{CaffeineMark_Code} benchmark. Results showed that NDroid-DVM and NDroid-ART could successfully track data leakage from multiple contexts. Compared with Qemu, NDroid were 10.7 and 10.1 times slower in terms of the DVM runtime and ART runtime, respectively. But it was still more efficient than DroidScope\cite{Kwong_C2012}, which was at least 11 times slowdown.

\textit{f. Context-aware information flow analysis on IoT apps.} Current IoT app's permission models in SmartThings platform have design flaws that may expose user's private data during inter-app cooperation. To solve these problems,  \cite{Jia_C2017} designed and implemented ContexIoT, a context-based permission system for IoT apps. ContexIoT supported fine-grained identification on a sensitive action. The context information and the data flows obtained by DTA was provided to the user to make a more informed decision at run time.

In their approach, the run-time context information included \textit{ID/PID}, \textit{UI Activity}, \textit{Control flow}, \textit{Runtime value} and \textit{Data flow}. ContexIoT firstly executed app patching during the installation time. Through the static analysis, ContexIoT efficiently identified all the potential sinks containing secure sensitive behaviours in the app code. Then ContexIoT efficiently patched the app with the context collection logic. Once the sensitive execution was triggered at run time, a context collection function collected the essential information from the environment. For the data flow context, ContexIoT performed DTA to track the \textit{data dependency} of sink-related variables. ContexIoT maintained a taint environment as a taint label for each variable. For the taint propagation, ContexIoT adopted the generic approach \cite{Chen_C2015} and supported handling Groovy-specific operations. ContexIoT also supports the detection of implicit flows. Finally, all contextual information was sent to the backend for requesting execution permission of the action.

The experiments were conducted on the SmartThings IDE simulator. The data sets included 25 own-made SmartApps and 283 commodity SmartApps. ContexIoT accurately patched all 72 potential sinks in 25 SmartApps, and all attacks could be successfully identified without any ambiguity. Besides, the app patching added 67.1 ms additional delay on average. 

\textit{g. Quarantined execution based on sandbox} For data flow protection in IoT apps, permission-based access control is not adequate. At the same time, static and dynamic taint analysis have their flaws on implicit flow detection \cite{Golam_C2013}, such as imprecision on flows and high extra overhead. To address these challenges, \cite{Fernandes_C2016} presented FlowFence, a system that enabled robust and efficient flow control in IoT apps. The key idea of FlowFence was based on its new information flow model, called as Opacified Computation. It supported quarantining parts of codes, \textit{i.e.}, Quarantined Modules (QM), for isolated execution in different sandboxes. FlowFence orchestrated execution by chaining QMs together via taint-tracked handles, \textit{i.e.}, opaque handles. And DTA was performed at the component level to protect data transmissions between different modules.

App's code was first partitioned into different parts/modules according to their functionalities. Then pieces of code operating on sensitive data were convert into QMs. After that, a set of QMs were chained together to achieve useful application. FlowFence protected transmissions of sensitive data among different QMs. It translated the return value of sensitive data to opaque handles. Opaque handles were immutable. They were dereferenced only by QMs when running inside a sandbox. Meanwhile, FlowFence tracked the data flows across different QMs by combing DTA at the component level, \textit{i.e.}, tainting on data sources, QM's executing sandboxes and opaque handles. Finally, when data flows were transmitted to the sinks, FlowFence prevented data leakage by enforcing flow policies which were declared in the app's manifest. Based on above design, FlowFence was implemented in an IoT hub running on Android OS .

They performed experiments on an LG Nexus 4 with Android 5.0. It took 2.7MB RAM on average and 49.5Mb RAM in total on 16 sandboxes. QM call latency was less than 100ms when there were four or more spare sandboxes. In addtition, they selected three typical IoT apps, including SmartLights, FaceDoor, and HeartRateMonitor. FlowFence could detect and eliminate those possible insecure flows effectively. The extra overheads were around 99 lines of code increment on per app and 4.9\% latency increment on regular function. 

\subsubsection{\textbf{Secure multi-execution}} Instead of tracking specific illegal flows in the application, SME provides a secure execution environment for the application to avoid data leakage \cite{Schmitz_C2018}. SME supports multiple executions on copies of application with different security levels. Flow control based on multiple executions aims to mitigate false alarms on illegal flow \cite{Rafnsson_C2016} \cite{Zanarini_C2013}. However, it also increases the overhead due to the repeated computations compared with the traditional approach. 

\textit{a. Context-sensitive DIFC enforcement on Android.} Decentralized IFC (DIFC) \cite{Myers_C1997} \cite{Nadkarni_C2013} \cite{Xu_C2015} is required for Android apps to specify control policy on their own data. Nevertheless, due to sharing state in memory and storage, former DIFC systems may cause the inefficiency on the app's execution, such as dangling state, dead lock, false-negative control, extra modifications on applications and so on. Therefore, Adwait \textit{et al.} \cite{Nadkarni_C2016} proposed a context-sensitive DIFC enforcement for Android applications via lazy polyinstantiation. They also designed the Domain Declassification mechanism for practical and secure network export.  

They proposed four design goals for their system (Weir), \textit{i.e., separation of shared state in memory (G1), separation of shared state on storage (G2), transparency  (G3)}, and \textit{secure and practical declassification for network export (G4)}. For G1, Weir used component polyinstantiation based on analyzing different labels of the instances, namely \textit{lazy polyinstantiation} to reuse a previously instantiated process. For G2, Weir separated shared state in the internal and external storage by using file-system polyinstantiation. It was based on a layered file system approach\cite{BROWN_code}. For an internet-driven environment, they adopted domain declassification for feasible and secure data export through the network, \textit{i.e., goal G4}. Besides, the component and storage polyinstantiation only required modifications on the system level, which satisfied G3. Based on the design, Weir modified the procedure of \textit{Android Activity Manager}, and was built on \textit{overlayFS} \cite{BROWN_code},  \textit{zygote}, and \textit{Linux security module}.

Security analysis on the polyinstantiation showed that polyinstantiation could be resistant to data leaks in floating labels. The experiments were conducted on 30 randomly picked top applications on Google Play. There was no crashes observed during the executions of two instances of the same activity. In addition, Weir incured less than 4 ms overhead to start components, 5.98 ms for the copy operation, and 2.02 ms for establishing a network connection. And the result showed that the start time increased linearly with the number of concurrent instances.  It all took 56 ms for the total 100 instances.

\textit{b. Secure multi-execution in Android.} SME was widely used in the browser \cite{Devriese_C2010} \cite{DeGroef_C2012} to prevent sensitive data leakage with soundness and precision guaranteed. Weir \cite{Nadkarni_C2016} used polyinstantiation technique but still did not construct a complete SME for Android apps. Chakraborty \textit{et al.} \cite{Chakraborty_C2019} proposed an IFC architecture named Ariel for Android based on the secure multi-execution of apps. Ariel realized the parallel execution of multiple same application instances. In addition, an I/O scheduler was used to control over data flows between application instances.

To realize SME in Android, they firstly implemented a multi-execution environment based on sandboxes by modifying the application start up process. They extended the default application start up process by modifying activity manager service (AMS). Then two or more different application processes could load with the same application code. Next, they defined a security lattice on application processes for enforcing IFC on apps. The lattice was derived from Android's permission system and had two security levels. After that, different types of components were labelled with security levels to securely process the inter-component communication (I/O operations in component). Finally, an I/O scheduling module was implemented inside the AMS. The module was responsible for intercepting and controlling all the intent-based communications routed through the AMS. 

They performed two case studies on the effectiveness of SME, including permission re-delegation attacks \cite{Felt_C2011} and data leakage from untrusted third-party code \cite{son_C2016}. Through the experiments, results showed that Ariel could prohibit these two attacks successfully.

\subsubsection{\textbf{Run-time enforcement based on type system.}} Type checking can occur statically (at compile time), dynamically (at run time), or as a hybrid of static and dynamic checking. Run-time IFC based on type system can be used to verify the features or properties which are difficult or impossible to verify statically, such as dynamically loaded classes, reflections, etc.

\textit{a. Real-time IFC on IoT apps based on type system.} DTA can not deal with implicit flow well. SME provides soundness but with high overhead, which is not practical for IoT apps. Thus, Bastys \textit{et al.} \cite{Bastys_C2018} developed a dynamic monitor framework, FlowIT,  to check IFTTT's (If This Then What) apps. The framework modeled applet reactivity and timing behaviour, and tracked the flows by a well-typed monitor. The monitor executed run-time detection on the illegal flows according to predefined typing rules. 

They first described the syntax of $filter$ functionality in IFTTT, which was built on an imperative core of JavaScript extended with APIs for sources and sinks. They assigned variables and sinks with specific security labels by defining typing environment. Security labels belonged to a two-level lattice. After that, the semantic rules were instrumented based on basic expression typing rules, including expression evaluation, command evaluation on skip, sink, assign, sequence, if and while operations. Besides,  new semantics for trigger-sensitive or trigger-insensitive applets were added to avoid leaking information from the termination channel. They also modeled applet termination by counting the steps in the filter semantics. If the execution of the filter code exceeded the steps allowed by the timeout, the monitor blocked the execution of the applet and performed no outputs. Therefore, they supported the protection on four types of flows, \textit{explicit flow, implicit flow, presence flow} and \textit{timing flow}. They implemented FlowIT based on the JSFlow \cite{hedin_C2014} tool.

For evaluations, the soundness of the designed monitor was proved. It showed that their monitor enforced projected noninterference. They also evaluated the monitor mechanism on a collection of 60 applets. For the filter code, they only observed one false positive with no false negatives.

\textit{b. Securing cross-app interactions in IoT Platforms.} Connected IoT apps can help users build powerful applications, but interactions between different apps may lead to sensitive information leakage. Some works use program analysis technique to reveal unexpected interference across applications \cite{Ding_C2018} \cite{Surbatovich_C2017}. Foundational questions remain largely unexplored, including the interaction semantics of apps, soundness of enforcement mechanisms, etc.  Balliu\textit{ et al.} \cite{Balliu_C2019} designed a calculus, CaITApp,  to model the run-time behavioural semantics of concurrently-executing  apps. They also used it to define required semantic policies of IoT apps in terms of security and safety context..

They first defined a calculus to specify and reason about the behavioural semantics of apps running on concurrent IoT platforms, called \textit{CaITApp}. Calculus provided syntax representation for IoT systems. Then they formalized a notion of \textit{safe cross-app interaction} based on a bi-simulation behavioural semantics. It is used to describe the semantics of cross-app interactions during run time. After that, they extended the framework to support implicit app interactions. This could reduce the challenge of false negatives and false positives. Finally, based on the above formal model, an extension of the flow-sensitive type system \cite{Hunt_C2006} was proposed for ensuring the confidentiality and integrity in the IoT app. Typing rules were defined according to the syntax of the designed calculus. The soundness of the designed type system was proved through contradiction methods. They also provided further examples based on real-world apps.

\section{Challenges and Future Works} \label{sec_discussion}
\begin{figure}[t]
\setlength{\abovecaptionskip}{0.2cm}
\centerline{\includegraphics[scale=.38]{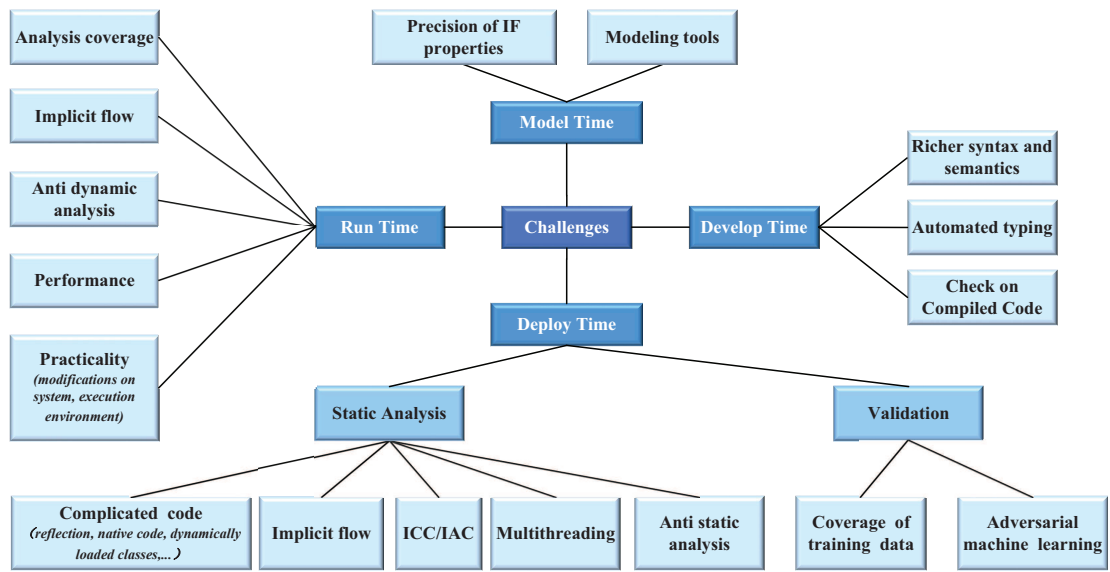}}
\caption{The Overview of Challenges in Different Phases}
\label{fig_challenge}
\vspace{-0.4cm}
\end{figure}

We have throughly reviewed the state-of-the-art systems/schemes of information-flow based defense mechanism for data leakage detection in smart applications. Nevertheless, it is a continuous work to protect users' private data because more complex smart apps applied in people's daily life. The involvement of AI-based applications and the rapid growth of IoT devices also increase the risk on data leakage. Many critical challenges are remained to be solved. In this section, we discuss the crucial research challenges and potential future directions in line with defensive chain. The overview of challenges in different phases is shown in Fig.\ref{fig_challenge}.

\subsection{Challenges and Future Directions on Model-time and Develop-time Flow Control}
\subsubsection{\textbf{Model-based information flow control}}
\textbf{Precise IF-properties modeling}. The precision and the coverage of IFC highly rely on the precision and granularity of the modeling process. There are still false negatives or positives due to the simplified behaviours and environments. Therefore, more specific IF properties modeling and verification still need to be considered. For example, \cite{Kartause_C2015} planed to support dynamic properties by allowing the user-defined privacy for different data and scenarios. They also planed to specify the behaviour of declassification methods in the model.

\textbf{Support of modeling tools}. For those IoT apps which are boomed in recent years, general and comprehensive modeling tools are still missing due to the filed-specific features. It poses a critical challenge to the implementation of model-based secure developing approaches. By combining with agile development for CPS system \cite{Geismann_C2018}, specific modeling and developing tools can be designed and developed for secure development on IoT applications.

\subsubsection{\textbf{Develop-time information flow control based on type system}} \textbf{Support on richer programming languages}. Most existing type systems are mainly built on an imperative language without richer syntax and semantics. This limits the formalization on real application and may miss some important features. Though \cite{Chen_C2018} introduced Android permission model into type system, some future extensions needed to be studied, including typing global variable\cite{Chen_C2018}, object-oriented feature \cite{Qi_C2004}, exceptions \cite{Barthe_C2006}, run-time permissions etc. 

\textbf{Automated typing algorithm}. Manual typing on each variable by the programmer is a time-expensive and effort-consuming work. Designing the automated typing algorithm should be one of the improvement on the security type system. Works on identifying sensitive sources and sinks based on ML and NLP \cite{Nan_C2015} \cite{Huang_C2015} \cite{Li_C2015} \cite{Nan_C2018} can be refereed.

\textbf{Check on compiled code.} Besides, develop-time analysis approaches mainly analyze the source code. Some researchers \cite{Chen_C2018} \cite{Ernst_C2014} tried to translate the typing on source code into typing on compiled code, \textit{e.g.}, Dalvik bytecode for Android apps. It can be adopted by those who can not obtain the source codes in most scenarios, such as app market owner.

\subsection{Challenges and Future Directions on Deploy-Time Flow Control}
\subsubsection{\textbf{Static information flow analysis}} Static analysis always suffers the precision and efficiency problems from the complicated codes, system-level mechanisms, dynamically loaded codes, etc. Although the surveyed works improve the static information flow analysis mechanisms from different aspects, there are still some challenges.

\textbf{Analysis on complicated code}. Static analysis on some complicated objects and functions are still not fully supported, mainly including string analysis \cite{Li_C2015} \cite{Wei_C2018} \cite{Celik_C2018}, reflection calls \cite{Li_C2015} \cite{Wei_C2018} \cite{Celik_C2018} \cite{Gordon_C2015} \cite{Backes_C2016} \cite{Calzavara_C2016}, native calls\cite{Li_C2015} \cite{Gordon_C2015} \cite{Backes_C2016} \cite{Calzavara_C2016}, and external libraries \cite{Gordon_C2015} \cite{Calzavara_C2016}. 

$\bullet$ String analysis: String analysis performs a critical part for analyzing reflection calls, ICC, \textit{etc}. \cite{Li_C2015} \cite{Wei_C2018} and \cite{Calzavara_C2016} partly supported simple string analysis. Nevertheless, precise analysis is difficult and expensive in both Java and binary code  \cite{LiDing_C2015}, which is left as the future work in most research works.  

$\bullet$ Reflection calls: Resolution on reflection methods is another challenge in static analysis. This is because reflective calls set their method through external configuration files, and Android does not support access to this kind of information statically during deploy time.  \cite{Li_C2015} resolved reflective calls only if their arguments are constant strings. \cite{Wei_C2018} cannot handle Java reflection without precise string analysis support. Celik's treatment on call by reflection increased the number of analyzed methods, which may lead to over-tainting \cite{Celik_C2018}. \cite{Gordon_C2015} and \cite{Calzavara_C2016} did not have fully handled of Java reflection, but they attempted to aggressively resolve reflection targets similar to \cite{Smaragdakis_C2015} and \cite{Livshits_C2005}.  Therefore, how to precisely resolve reflection methods during tracking flows is still a continuous and critical work in the future.

$\bullet$ Native calls: Before Wei's work \cite{Wei_C2018}, most information flow tracking tools \cite{Li_C2015} \cite{Gordon_C2015} \cite{Backes_C2016} \cite{Calzavara_C2016} do not support the analysis on native methods. \cite{Wei_C2018} was implemented on top of AmAndroid \cite{Wei_C2014}. It is still a challenge to integrate the inter-language analysis into other static analysis tools. For example, \cite{Calzavara_C2016} planed to support the analysis on native code based on SMT solving, which has been successfully applied in the C program's analysis.

$\bullet$ External libraries: Due to heavy overhead on the application-related external libraries provided by application framework, many tools adopted some shortcuts to tackle it, such as assuming on trusted framework and libraries \cite{Gordon_C2015}, \cite{Celik_C2018} , or treating the libraries as a black box and performing conservative analysis on it \cite{Li_C2015} \cite{Calzavara_C2016}. Hence, it is a challenging work that combines the analysis of external libraries with limited efforts in the future.

Some features are still not supported for different analysis approaches, including multithreading analysis, implicit flow detection, ICC analysis, \textit{etc}. 1) Multithreading introduces non-deterministic and probabilistic execution path, which makes static analysis more difficult \cite{Arzt_C2014}. Some works treated multithreading as multiple threads executed in sequential but arbitrary order. However, fine-grained analysis on each possible path may result in \textit{Path$\&$State} explosion problems \cite{Wei_C2018}. \cite{Calzavara_C2016} performed conservative analysis trading precision for efficiency and soundness. 2) Some analysis approaches do not support detection on implicit flow \cite{Celik_C2018} \cite{Gordon_C2015} \cite{Calzavara_C2016} and flows through ICC \cite{Calzavara_C2016}, which can be further extended in the future. 3) platform-specific analysis approaches, such as Droidsafe built on Android 4.4.3 \cite{Gordon_C2015} and SAINT built on SmartThings programming platform \cite{Celik_C2018}, still need to be improved when there are changes on the platform (\textit{e.g.} upgrades on execution environments, deployment on a new platform). 

\textbf{Anti-analysis attack}. Most static analysis approaches can not resist side channel and code-obfuscation attack. Dynamic loading classes can not be analyzed either. Other techniques such as, the countermeasure against side channel attack\cite{Barthe_C2016}, anti-obfuscation techniques \cite{Rasthofer_C2016}, and dynamic analysis \cite{You_J2017}, can be considered to solve these challenges.

\subsubsection{\textbf{Validation on flow's legality}} Semantic-based validations on detected flow provides an efficient way to automatically detect leakage, but limitations and attacks on machine learning pose significant challenges on this type of approaches.

\textbf{Coverage of training data}. The knowledge about flows limits the precision of identification. Different works are based on different knowledge/dataset, such as similar application's policies \cite{Lu_C2015}, detected malware \cite{Vitalii_C2015}, privacy policies \cite{Nan_C2018},  key words\cite{Huang_C2015}, application context \cite{Pan_C2018}. Integrating pieces of knowledge to improve the precision has been considered as future work in many works, \textit{e.g.}, app descriptions in \cite{Vitalii_C2015}, context information in \cite{Huang_C2015}, app UI in \cite{Pan_C2018}.

\textbf{Adversarial machine learning}. These identification techniques are prone to polluting attacks due to the limitation of protection on analysis data. Peer voting mechanism \cite{Lu_C2015} could be disturbed by polluting the peer apps. MUDFLOW \cite{Vitalii_C2015} relied on the features of the detected flows, in which deficiencies of training data would impair the classification. Besides, clickjacking and UI redress attacks in \cite{Pan_C2018} and code obfuscation in \cite{Nan_C2018} can pollute the real-world data, which also could lead to inaccurate detection. Therefore, the authenticity and integrity of the data should be ensured during the analysis.

Furthermore, detected flows are obtained from the static information flow analysis techniques. These approaches also inherited the drawbacks of the static analysis that they adopted.

\subsection{Challenges and Future Directions on Run-Time Flow Control}
\subsubsection{\textbf{Dynamic taint analysis}}Due to the limitations on dynamic analysis, there are still several challenges existing in these approaches. For some challenges, it may need to combine with static analysis techniques to tackle with.

\textbf{Analysis on implicit flow}. Implicit flow detection is still not supported by most dynamic taint analysis \cite{Sun_C2016} \cite{Wu_J2018} \cite{Xue_J2019} \cite{Yoon_C2016}. \cite{You_J2017} applied static analysis for capturing strict control dependence caused by equivalence. But the analysis on other control dependence($>$, $<$, etc.) needed the constraint solvers. Besides, nested control dependency has not been supported in \cite{You_J2017}. \cite{You_J2017} planed to overcome this challenge by maintaining a stack of implicit flow labels, but it may cause a high-performance impact.

\textbf{Analysis coverage}. Path coverage is another major challenge due to the natures of dynamic analysis. It is required to manually trigger the behaviours \cite{Sun_C2016}, which consumes a lot of human effort. Equipping with advanced input generation system \cite{Machiry_C2013} for high-coverage analysis is one of the interesting directions in the future \cite{wong_C2016} \cite{Xue_J2019}.  

\textbf{Anti-analysis attack}. There are some anti-analysis techniques that aim to evade the analysis, such as identifying running environment \cite{petsas_C2014}, code obfuscation \cite{Yoon_C2016}. By combining with virtualization technology supported by CPUs (\textit{e.g.}, Trustzone in ARM \cite{Trustzone_Code}) can be a promising approach to tackle the system identification attack \cite{Xue_J2019}. Besides, handling the obfuscated code effectively are also considered in the future \cite{Yoon_C2016}.

\textbf{Overtainting}. Overtainting is one of the most common challenges in taint analysis techniques. It may lead to poison-pill attacks \cite{Hritcu_C2013}. In this case, the taint bound mechanism \cite{Fernandes_C2016} needs to be studied for preventing overtainting. 

\textbf{Validation on flow's legality}. It is hard for the user to make right decision on each detected flow without domain knowledge. Hence, it can be improved by providing more auxiliary information on flows, such as frequency of a granted permission \cite{Almuhimedi_C2015}, to assist users in evaluating flows. Meanwhile, automatically verification mechanism based on experts' knowledge is another interesting direction. 

\textbf{Practicality of dynamic analysis}. The practicality of different dynamic analysis approaches is also impacted by \textit{the required modification level of the system}, \textit{specific execution environment/platform} \textit{etc}. 

$\bullet$ Modification on System: Previous work \cite{Sun_C2016} needs to modify and re-compile the Android framework code, which requires the root privilege. In \cite{You_J2017}, dynamically loaded codes need to be obtained and instrumented in advance (off-line), which may affect the integrity checking. Reference hijacking in \cite{You_J2017} requires to repackage the target application, which suffered from the re-signing issues. Therefore, dynamic tainting and tracking techniques are required to improve further with fewer modifications and privilege requirements, \textit{e.g.}, execution-time instrumentation \cite{You_J2017} and boxify (sandbox) techniques \cite{Michael_C2015}. 

$\bullet$ Execution environment/ platform: Dynamic analysis approaches are built on specific execution environment/platforms, such as \cite{Sun_C2016} on 32-bit ARM-based device, \cite{Xue_J2019} on Android emulator, \cite{Wu_J2018} on DVM, \cite{Jia_C2017} on the trustworthy and uncompromised platform, \textit{etc}. Supporting newly updated environment/platforms with fewer security assumptions is a continuous work for DTA approaches. 

In addition, other features also can be improved in the future, including supports on more types of apps \cite{Xue_J2019}, performance improvement\cite{Yoon_C2016}, validation model and policy \cite{Wu_J2018} \cite{Jia_C2017}, declassification\cite{You_J2017}, \textit{etc}. 

\subsubsection{\textbf{Secure multi execution}}SME provides a sound way to protect sensitive information flow with a secure paralleling and isolated execution environment. However, it also faces the following challenge.

\textbf{Performance of SME}. The performance is impacted with overhead caused by multi-execution instances, especially for multi-level security requirements\cite{Nadkarni_C2016}. On the other hand, attackers may exploit the Denial of Service attack on specific application components. Attackers could start a vast number of the instances to break down the SME \cite{Chakraborty_C2019}. Therefore, performance improvement and anti-resource-consuming mechanisms should be studied and adopted in future. Besides, research on SME in mobile devices is at an early phase while none is adopted in IoT devices. More features on SME should be improved in the future, including more complex security policies, declassification policies, inter-component communications \textit{etc}.

\subsubsection{\textbf{Run-time check based on type system}}
\textbf{Support richer language and run-time behaviours}. Similar to develop-time analysis , existing works focused on simplified language and execution behaviours for sound verification on core functions. Therefore, it is a continuous work to improve the type system to support richer language and more run-time behaviours in different platforms.\cite{Balliu_C2019}.

\section{Conclusion} \label{sec_conclusion}
In this survey, we present a review of IFC techniques used for data leakage detection in smart applications. We firstly summarize the research methodology as a defensive chain which integrates different IFC techniques covering all critical phases of the application's life cycle, including model-time, develop-time, deploy-time and run-time flow control. According to the defensive chain, a detailed literature review is conducted based on recent research efforts on IFC used in Android and IoT applications. Approaches in each phase are categorized based on different analysis techniques and protection mechanisms. Furthermore, in-depth analysis and comparisons are made for different IFC approaches in each phase. It provides a valuable reference on applying appropriate IFC technique according to different security requirements, such as precision, performance, execution environment, application's life phase, \textit{etc}. In the end, existing challenges for IFC techniques are discussed, along with future directions being elaborated. 

With the emerging smart applications in various mobile and IoT platform, data leakage detection based on IFC techniques have aroused more and more concerns from both the academic and industrial field. The survey can provide valuable insights for further research of software security community.




\bibliographystyle{ACM-Reference-Format}
\bibliography{SurveyOnED}


\end{document}